\documentclass[twocolumn,prb]{revtex4}
\usepackage{graphicx}

\newcommand{\Tr}{{\rm Tr}} 
 
\renewcommand{\Im}{{\rm Im}} 

\begin{document}
\title{Spin correlations in the algebraic spin liquid - implications for
high $T_{c}$ superconductors}
\date{\today}
\author{Walter Rantner}
\author{Xiao-Gang Wen}
\affiliation{Dept.\ of Physics, Massachusetts Institute of Technology,
Cambridge, Massachusetts 02139}
\begin{abstract}
We propose that underdoped high $T_c$ superconductors are described by an
algebraic spin liquid (ASL) at high energies, which undergoes a spin-charge
recombination transition at low energies. The spin correlation in the ASL is
calculated via its effective theory - a system of massless Dirac fermions
coupled to a $U(1)$ gauge field.  We find that without fine tuning any
parameters the gauge interaction strongly enhances the staggered spin
correlation even in the presence of a large single particle pseudo-gap.  This
allows us to show that the ASL plus spin-charge recombination picture can
explain many highly unusual properties of underdoped high $T_c$
superconductors.
\newline
\newline
\end{abstract}
\pacs{PACS numbers: 75.10.-b, 74.72.-h, 76.20.+q}

\maketitle

\section{Introduction}

One of the intriguing questions about the cuprate superconductors - to many
possibly the key to understanding superconductivity - is the role played by
spin correlations in these materials.  By now it is well established that the
insulating parent compound of the copper oxide superconductors is well
described as a two dimensional Heisenberg Antiferromagnet in the temperature
regime above the three dimensional Neel ordering temperature ($T_{N} \sim
300K$). Between $T_{N}$ and a temperature of about 600 K the materials seem to follow the renormalized classical results of Chakravarty, Halperin
and Nelson \cite{CHN}. At higher temperatures the system appears to cross over
to quantum scaling behavior with a dynamical exponent $z = 1$
\cite{SlichterLAS}.  On doping with holes, away from stoichiometry, these
insulating compounds develop into high $T_{c}$ superconductors even for very
low hole concentration of order 5\%. For such low doping the low energy
Hilbert space should still be dominated by spin fluctuations - as a memory
effect of the undoped parent compound.  The question that needs to be
addressed is the peculiar interplay of short range antiferromagnetic
correlations as a remnant of the ordered Neel state at zero doping competing
with spin singlet formation present in the superconducting state.
Depending on the importance attached to those two spin correlations we can
distinguish two schools of thought that have strongly influenced the high
$T_{c}$ research.

The first takes the {\emph strong} antiferromagnetic (AF) spin fluctuations
seriously as providing the ``mechanism'' for superconductivity.  Assuming the
normal state is a Fermi liquid, the antiferromagnetic spin fluctuations were
shown to lead to a $d$-wave superconducting state.\cite{Scalapino} The spin
fluctuations can be modeled phenomenologically a la Millis, Monien and Pines
\cite{MMP} in the context of a nearly antiferromagnetic Fermi liquid (NAFL).
The main degrees of freedom are spin and charge carrying quasiparticles
coupled to their own spin fluctuations. The strength of the spin fluctuations
is modeled by a susceptibility which is strongly peaked near the commensurate
antiferromagnetic wavevector $\vec{Q} = (\pi/a,\pi/a)$. This strong
enhancement is achieved via an RPA in the form
\begin{displaymath}
\chi(\mathbf{q},\omega) =
\frac{\chi_{0}(\mathbf{q},\omega)}{
1-J(\mathbf{q},T)\chi_{0}(\mathbf{q},\omega)}
\end{displaymath}
as proposed by Monthoux and Pines ($\chi_{0}(\mathbf{q},\omega)$ is calculated
by taking spin fluctuations mediated interactions between the quasiparticles
into account).  In a series of papers\cite{MP} Monthoux and Pines have given
what they call a ``proof of concept'' for spin-fluctuation driven pairing for
the cuprates near optimal doping. They study the competition of spin-density
wave (SDW) order with superconducting pairing by solving two-dimensional
Eliashberg equations for spin-fluctuation induced interactions between the
quasiparticles. These approaches could account for both the d-wave pairing
symmetry and the high $T_{c}$'s seen in the cuprates.  One interesting
question is of course the interplay of SDW and superconductivity as
instabilities of the NAFL. Monthoux and Pines concluded that superconductivity
always prevails as the dominant instability. With the realization that the
quasiparticles are not too close to a SDW instability they however needed a
strong fine tuning of their Stoner enhancement factor $J(\mathbf{q})$ as a
function of both momentum dependence (strongly peaked at $\mathbf{Q} =
(\pi/a,\pi/a)$) and strength ($J(\mathbf{Q})\chi_{0}(\mathbf{Q},0) \sim 0.97$)
to account in a ``self-consistent'' manner for their empirical input spin
spectrum.  

The discussion so far was centered on the optimally doped case, in this paper
we shall however be concerned with the underdoped  cuprates where the spin
behavior shows the peculiar competition of antiferromagnetic order and singlet
formation in a particularly striking way as is evidenced by various
spin-pseudo-gaps
seen in NMR and Neutron scattering. Let us briefly mention the approaches to
this regime which have emphasized the antiferromagnetic order and the
concomitant spin wave excitations as the dominant degrees of freedom.  In an
extensive study of the $2d$ antiferromagnetic Heisenberg model Chubukov,
Sachdev and Ye \cite{CSY} have identified the spin correlations in the quantum
critical regime at the transition from Neel order to paramagnetism.  In their
paper they also argue for the appearance of quantum critical scaling away from
zero doping by incorporating the effect of the doped charge degrees of freedom
into a reduction of the spin stiffness (finite in the ordered Neel state of
the 2d Heisenberg model) Sokol and Pines\cite{SokolPines} have similarly
argued for quantum critical behavior and its cross over to a quantum
disordered regime driven by the frustration of antiferromagnetism via the hole
motion as the physics behind the strange spin correlations in the
pseudo-gap regime.  They postulate a susceptibility which is dominated by
the effect of a finite coherence length in the quantum disordered regime and
the appearance of the corresponding \emph{full} energy gap for spin
excitations.

As we shall argue below this phenomenology qualitatively agrees with what we
obtain within the slave boson approach to the $tJ$-model, where however in
contrast to the above, spin singlet formation (or more precisely, the
formation of the d-wave/staggered-flux phase -- a state with a particular
quantum order U1Cn01n\cite{Wqos,Wqo}) is the driving force behind all of the
strange phenomenology in the underdoped samples.  The slave boson approach is
a microscopic theory for high $T_c$ superconductors which takes the on-site
repulsion $U$ and the resulting charge gap ($\sim 2$eV) as the largest scale
in the theory. It naturally reproduces the Mott
insulator phase at half filling and low charge carrier density at low doping.
Furthermore this framework  allows us to calculate various line-shapes when the
AF-correlation is short ranged. The  line-shapes can be quite different from
those obtained from phenomenological models.  

The slave boson approach is strongly tied to the strong coupling phenomenology
incorporated in the Hubbard or $tJ$-model and was guided by Anderson's
\cite{RVB} exciting proposal of a spin liquid as a realization of the strongly
correlated Mott insulator in the parent compound of the high $T_c$
superconductors. More specifically, Anderson proposed that the cuprate physics
could be understood in terms of doping holes into a state which consists of
preformed spin singlet dimers on nearest neighbor bonds whose quantum
fluctuations lead to uniform strength AF-correlations on all nearest neighbor
bonds (termed resonating valence bond (RVB) state). The RVB state also has
short range antiferromagnetic spin correlations (no true long range order) due
to the singlet formation and hence comprises a spin liquid phase.  However,
due to the success of the two dimensional Heisenberg antiferromagnet in
describing the insulating parent compound of the cuprates this idea of a spin
liquid ground state was partly rejected.  

Nevertheless the RVB picture has a lot of appeal. Let us mention briefly the
rational behind the opinion that the spin singlet formation dominates
antiferromagnetic correlations in the underdoped cuprates.  The slave boson
approach treats the underdoped cuprates as doped Mott insulators - the key
experimental fact of high $T_c$ superconductors.  It also naturally
incorporates the spin singlet formation and predicted, ahead of experiments, the
pseudo-gap metallic phase\cite{AM8874} 
in the underdoped cuprates and the superconducting
d-wave order (preferred over s-wave by strong Coulomb correlations present in
the Mott insulator).\cite{Kotliar} The pseudo-gap metallic phase is a new state
of matter and it is very rare in the history of condensed matter physics that
a new state of  matter was predicted \emph{before} its experimental observation.
An important effect of the preformed spin singlets, present in the RVB
picture, is the fact that the spin on the doped holes can become an excitation
on top of the RVB collective state whereas the charge remains tied to the
empty site.  This led to the notion of \emph{separate} spin and charge
carrying excitations called spinons and holons respectively.
Superconductivity then appears as a consequence of establishing coherence
among the doped holes which naturally incorporates a low (proportional to
$x$ the hole doping concentration) superfluid density present in the
underdoped cuprates.  This picture of the spin fluctuations is quite different
from the NAFL where (1) a metallic state with no pseudo-gap is assumed (which
does not apply to the underdoped regime) and (2) strong antiferromagnetic spin
fluctuations couple to spin and charge carrying quasiparticles. It is the
exchange of those antiferromagnetic fluctuations which gives rise to the
pairing necessary for superconductivity.  

However, underdoped cuprates have a very puzzling property which seems hard to
explain using the spin liquid approach (or other approaches).  As the doping
is lowered, both the pseudo-gap and the AF correlation in the normal state
increase.  Naively, one expects the pseudo-gap and the AF correlations to work
against each other. That is the larger the pseudo-gap, the stronger the
spin-singlet formation and the weaker the AF correlations.  One result coming
out of the research reported here is a connection of the quantum scaling
regime of the undoped compounds at high temperature with Anderson's RVB
picture via the algebraic spin liquid (ASL) phase. 
We find that the ASL can explain both the pseudo-gap and the strong AF
correlations in the underdoped high $T_c$ superconductors. Due to the $U(1)$
gauge fluctuations in the ASL phase, the AF spin fluctuations in the ASL are
as strong as those of a nested Fermi surface, despite the pseudo-gap. The
doped holes suppress the $U(1)$ gauge fluctuations which explains the decrease
of the AF spin fluctuations with increasing doping even though the pseudo-gap
is smaller at higher doping.  We would like to stress that the ASL approach to
the AF fluctuations is distinct from RPA-like approaches.  Due to the small
density of states in the pseudo-gap phase, one needs to fine tune a large
coupling constant within the RPA to fit the experimental data for underdoped
samples. Such a fine tuning is not necessary in the ASL approach.
Thus the $U(1)$ gauge structure (and the associated spin-charge separation) in
the ASL lie at the heart of the strange phenomenology of the underdoped
cuprates in the so called pseudo-gap phase \cite{ASL}.


The starting point of the slave boson approach is a microscopic lattice model.
Among the popular models, the $tJ$-$GtJ$ (\emph{G}eneralized-$tJ$) model
seems particularly promising as a description for the low doping regime where
the competition between delocalization energy $t$ and the spin fluctuations
$J$ becomes manifest.
\begin{equation}\label{GtJ}
H = \sum_{(ij)}\big[ J(\vec{S}_{i}\cdot\vec{S}_{j} - \frac{1}{4}n_{i}n_{j})
- t(c^{\dag}_{\alpha i}c_{\alpha j} + h.c.)\big] + ...
\end{equation}
The form of the $tJ$ Hamiltonian can be justified starting from the Hubbard
model in the limit of strong on-site Coulomb repulsion energy which in the
cuprates leads to an insulating charge transfer gap of $2eV$. As the Coulomb
energy is the largest energy scale in the problem it is natural to treat the
kinetic energy as a perturbation which to lowest order in $t^2$ leads to the
$tJ$ Hamiltonian with the single site Hilbert space restricted to the three
states spin up/down and empty.  For zero doping the $tJ$ Hamiltonian reduces
to the Heisenberg model with antiferromagnetic exchange coupling (virtual
hopping on top of strong onsite  Coulomb repulsion naturally leads to
antiferromagnetic exchange).

The organization of the paper is as follows; In the next two  sections
(\ref{ASL},\ref{susceptibility}) we review the ideas underlying the ASL and
discuss the spin correlations in this phase sketching the calculations
involved (details can be found in the appendix).
Following this we give a brief discussion on how the recently introduced concept of quantum order \cite{Wqos,Wqo} affects the current results (\ref{ASLPSG}).
 Section (\ref{fate})
addresses the problem of what happens to the ASL physics at low energies where
the interactions between spinons and holons become strong.
This will be followed up by a discussion of  possible implications for the
cuprates and a short comparison with other approaches to the same region of
the phase diagram (\ref{Discussion}). The final section (\ref{Conclusion})
will summarize the main results and discuss open problems for further study.

\section{$SU(2)$ Slave Boson Approach and Algebraic Spin Liquid}\label{ASL}

Taking the charge gap as the largest energy scale, we have argued that the
natural starting point to the description of underdoped cuprates is a strongly
correlated model.  The question of spin correlations in the cuprates ties into
the question of how to reconcile local moment magnetism with itinerant
electron spin density fluctuations which is a longstanding problem in
condensed matter physics.  There seems to be consensus in the community on how
best to describe the two extreme limits of the cuprate phase diagram - the
Mott insulator at zero doping as a Heisenberg antiferromagnet and the heavily
doped regime in terms of itinerant electron magnetism.  Of course as always
the most interesting regime is the one lying in between the two limits which
shows the antiferromagnetic-singlet dichotomy described above.  For the
lightly doped regime one could argue that clearly a perturbation about the
Neel ordered state with frustration due to holes, a la Sokol and
Pines\cite{SokolPines}, makes the most sense. However, this approach does not
specify the nature of the quantum spin liquid in the disordered phase.
Understanding the properties of the quantum spin liquid is very important
since it is the quantum spin liquid that controls the pseudo-gap metallic
state in the underdoped cuprates.  Another drawback with the NAFL approach to
the spin and charge fluctuations is the fact that it cannot explain naturally
the small charge carrier density without breaking translation symmetry. 

The picture of spin fluctuations followed here is the one of preformed
singlets (RVB) which incorporates the spin correlations in the spin disordered
phase (such as the superconducting state and the pseudo-gap normal state) in a
natural way by paying the prize of throwing away a lot of antiferromagnetic
correlations.  The aim of the current paper is to show a way of reconciling
the singlet formation with the strong dynamic antiferromagnetic correlations
seen in NMR and Inelastic Neutron Scattering (INS) with the help of a gauge
theory based on a slave boson decoupling of the $tJ$-model.  

As we discussed briefly above, the (generalized) $tJ$-model can be shown to be
an excellent description of the low energy physics embodied in the Hubbard
model under the condition that the single site Hilbert space is constraint
such that double occupation is forbidden.  On an operative level this problem
can be attacked by introducing slave particles that change the constraint
$c^{\dag}_{\sigma}c{_\sigma}(i) \leq 1$ (where $c_{\sigma}(i)$ is the physical
electron destruction operator of spin $\sigma$ on site $i$) into
$\Sigma_{\sigma}f^{\dag}_{\sigma}f_{\sigma}(i) + b^{\dag}b(i) = 1$ where the
$b(i)$ is a bosonic operator - the holon - that keeps track of the empty sites
and carries the charge of the physical hole, whereas the spin of the hole is
carried by $f_{\sigma}$ - the fermionic spinon. This is achieved by writing
$c_{\sigma} = f_{\sigma}b^{\dag}$ and should be read as an equality in the
constraint Hilbert space. With the introduction of the slave bosons we have
however introduced a redundancy in the description of our physical problem
which is a $U(1)$ phase rotation of $f$ and $b$ that leaves the physical
electron operator invariant. As the system evolves under the $tJ$ Hamiltonian
(\ref{GtJ}) this phase will strongly fluctuate as a function of space and
time.  It is this phase degree of freedom that corresponds to $U(1)$ gauge
fluctuations
and makes the description of the constraint problem of spin + hopping in terms
of fermionic and bosonic operators possible.  The above program can be
summarized under the name of $U(1)$ slave boson theory and was implemented
widely in the early days of high $T_{c}$
research\cite{U1}. Early on it was also
realized\cite{Affleck_ZHA} that the Heisenberg model in the fermionic
representation has an extra $SU(2)$ invariance organizing the spinons into
doublets
\begin{eqnarray}
\psi_{\uparrow i} = {f_{\uparrow i} \choose f^{\dag}_{\downarrow i}}, \quad
\psi_{\downarrow i} = {f_{\downarrow i} \choose -f^{\dag}_{\uparrow i}},
\quad
\end{eqnarray}
In the usual $U(1)$ slave boson approach this invariance was lost on
introducing holes. More recently Wen and Lee introduced a slave boson
formulation\cite{WenLee,fourauthor} which maintains this $SU(2)$ structure
away from half filling.  This was achieved  via the introduction of slave
boson doublets and representing the physical electron operator as
\begin{eqnarray}\label{SU2rep1}
c_{\uparrow i}=\frac{1}{\sqrt{2}}h^{\dag}_{i}\psi_{\uparrow i} =
\frac{1}{\sqrt{2}}\big(b^{\dag}_{1i}f_{\uparrow
i}+b^{\dag}_{2i}f^{\dag}_{\downarrow i}\big) \\ \nonumber
c_{\downarrow i}=\frac{1}{\sqrt{2}}h^{\dag}_{i}\psi_{\downarrow i} =
\frac{1}{\sqrt{2}}\big(b^{\dag}_{1i}f_{\downarrow
i}-b^{\dag}_{2i}f^{\dag}_{\uparrow i}\big) \\ \nonumber
\end{eqnarray}
with
\begin{displaymath}
h_{i} = {b_{1i} \choose b_{2i}} \nonumber
\end{displaymath}
the doublet of bosonic fields keeping track of the doped holes.

The slave boson approach to the $tJ$ Hamiltonian then is to use this
decoupling and perform a mean-field analysis \cite{Kotliar,U1,WenLee}.  This
has led to a mean-field phase diagram as a function of hole doping $x$ and
temperature $T$ which is in qualitative agreement with the phase diagram of
the cuprates.  Within the mean-field description however the gauge freedom is
treated only on the average - i.e. replaced by a static configuration of
phases which in turn determine the band-structure of the spinons and holons.
However as is usually the case with the identification of phases via a
mean-field decoupling  we need to consider fluctuations about the mean-fields
to determine their stability. In particular the gauge field - unconstrained by
any dynamics - will affect the physical properties in each phase drastically.

In the present paper we are concerned with the pseudo-gap regime of the
cuprates which - within the $U(1)$ formulation - was identified as the d-wave
paired state for the spinons.  Within the $SU(2)$ approach this phase can also
be described as the so called staggered-Flux (sF) phase without explicit
fermion pairing.  As it turns out (see \cite{fourauthor} for more details) the
$SU(2)$ formulation allows for a more straightforward identification of
low-lying (massless) gauge modes and it was shown that the sF phase breaks the
$SU(2)$ gauge structure down to $U(1)$. This massless $U(1)$ mode was missed
in the early discussions (within the $U(1)$ slave boson approach) on the
pseudo-gap phase and plays a crucial role in
the pseudo-gap phase of the underdoped cuprates.  It is responsible for the
emergence of the ASL.
In the next section we shall discuss the physical spin correlations in the
ASL.

Before closing this section, we would like to remark that the ASL (the sF phase) is only
one of many possible symmetric spin liquids.  All symmetric spin liquids have
the same symmetry, and hencewe cannot distinguish different symmetric spin liquids
by their symmetries. The concept of quantum order was
introduced\cite{Wqos,Wqo} to distinguish the different internal structures
present in the symmetric spin liquids.  For the symmetric spin liquids
constructed within the $SU(2)$ slave boson approach, one can use the
projective symmetry group (PSG) to characterize their different quantum
orders.  One finds that the sF phase is described by the particular PSG with the
name U1Cn01n and it is one of an infinite number of possible symmetric $U(1)$
spin liquids. Thus the sF phase can be more accurately called U1Cn01n phase.
Quantum order and its PSG characterization is a very important concept for our
discussion (See section \ref{ASLPSG}).

\section{Spin susceptibility}\label{susceptibility}

Our starting point is the sF (or U1Cn01n) state in the $SU(2)$ mean-field
phase diagram where the effective degrees of freedom are spinons and holons
coupled to a massless $U(1)$ gauge field.  In order to analyze this problem we
have mapped the lattice effective theory for the sF state (at zero doping)
onto a continuum theory of massless Dirac spinors coupled to a gauge
field,\cite{MarstonAffleck} whose Euclidean action reads
\begin{eqnarray}\label{QED3a}
S&=&\int d^{3}x \sum_\mu \sum_{\sigma=1}^N\bar{\Psi}_{\sigma}
v_{\sigma,\mu} (\partial_{\mu}-ia_{\mu})\gamma_{\mu}\Psi_{\sigma}
\end{eqnarray}
where $v_{\sigma, 0}=1$ and $N=2$, but in the following we will treat $N$ as
an arbitrary integer.  In general $v_{\sigma, 1}\neq v_{\sigma,2}$. However,
for simplicity we will assume $v_{\sigma,i}=1$ here.  The Fermi field
$\Psi_{\sigma}$ is a $4\times 1$ spinor which describes lattice spinons with
momenta near $(\pm \pi/2, \pm \pi/2)$.  The $4\times4$ $\gamma_{\mu}$ matrices
form a representation of the Dirac algebra
$\{\gamma_{\mu},\gamma_{\nu}\}=2\delta_{\mu\nu}$ ($\mu,\nu = 0,1,2$) and are taken to be
\begin{eqnarray}
\gamma_{0}&=&\pmatrix{\sigma_{3}&0\cr0&-\sigma_{3}\cr}, \quad
\gamma_{1}=\pmatrix{\sigma_{2}&0\cr0&-\sigma_{2}\cr}, \\
\gamma_{2}&=&\pmatrix{\sigma_{1}&0\cr0&-\sigma_{1}\cr}
\end{eqnarray}
with $\sigma_{\mu}$ the Pauli matrices.
Finally note that $\bar{\Psi}_{\sigma} \equiv \Psi^{\dag}_{\sigma}
\gamma_{0}$. 
The dynamics for the $U(1)$ gauge field arises solely due to the screening by
bosons and fermions, both of which carry gauge charge. In the low doping
limit, however, we will only include the screening by the fermion
fields,\cite{KimLee} which yields
\begin{eqnarray}
\cal Z&=&\int Da_{\mu}\exp\Big( -\frac{1}{2}\int\frac{d^3q}{(2\pi)^3}a_{\mu}
(\vec{q})\Pi_{\mu\nu}a_{\nu}(-\vec{q})\Big) \nonumber \\
\Pi_{\mu\nu}&=&\frac{N}{8}\sqrt{\vec{q}^2}\Big(\delta_{\mu\nu}
- \frac{q_{\mu}q_{\nu}}{\vec{q}^{2}}\Big)
\label{Pi}
\end{eqnarray}
By simple power counting we can see that the above polarization makes the gauge
field a marginal perturbation at the free spinon fixed point.  Importantly
however we should note that since the conserved current (that couples to
$a_\mu$) cannot have any anomalous dimension, this interaction is an
\emph{exact} marginal perturbation protected by current conservation.

\subsection{Uniform correlations}  

Let us now discuss how the gauge fluctuations affect spin correlations near momentum transfer $\mathbf{q} = (0,0)$.  
The expression for the uniform spin correlation reads (see Appendix
\ref{appA})
\begin{eqnarray}
\langle S^{+}_{u}(x) S^{-}_{u}(0) \rangle & = &\int \frac{d\vec{q}}{(2
\pi)^3} e^{i\vec{q}\cdot \vec{x}}\langle S^{+}_{u}(\vec{q})
S^{-}_{u}(-\vec{q}) \rangle \\ \nonumber
\langle S^{+}_{u}(\vec{q}) S^{-}_{u}(-\vec{q}) \rangle & = & -\frac{1}{4}
\int\frac{d\vec{p}}{(2\pi)^3} \Tr\big[\gamma_{0} G(\vec{p}) \gamma_{0}
G(\vec{p}-\vec{q}) \big]
\end{eqnarray}
where $\langle \cdots \rangle$ denotes the expectation value with respect to
theory (\ref{QED3a}).  From this expression we see that the uniform spin
correlation is proportional to $\Pi_{00}$, the polarization operator of the
spinons. Hence it cannot be strongly affected by the massless gauge field as
we argued above via current conservation. This was shown in an explicit
calculation by Chen, Fisher and Wu in the context of the FQH effect
\cite{CFW}.  Contrasting to this as was suggested by Kim and Lee \cite{KimLee}
we would expect the gauge fluctuations to strongly affect the staggered spin
correlations which are not protected by current conservation and restore
antiferromagnetic correlations\cite{Ivanov} which have been largely lost in
the mean-field singlet state as we shall see next.

\subsection{Staggered correlations}

In the following we will present fluctuation corrections to the
antiferromagnetic spin-spin correlations at order $1/N$.  The expression for
the staggered correlation is obtained in the form (see Appendix \ref{appA})
\begin{eqnarray}
\langle S^{+}_{s}(\vec{q}) S^{-}_{s}(-\vec{q}) \rangle & = & -\frac{1}{4}
\int\frac{d\vec{p}}{(2\pi)^3} \Tr\big[\openone G(\vec{p}) \openone
G(\vec{p}-\vec{q}) \big]
\end{eqnarray}
At the mean-field level this is simply
\begin{displaymath}
\langle S^{+}_{s}(\vec{q}) S^{-}_{s}(-\vec{q}) \rangle_{0} =  - \frac{1}{4}
\int\frac{d\vec{p}}{(2\pi)^3} \Tr\big[\openone G_{0}(\vec{p}) \openone
G_{0}(\vec{p}-\vec{q}) \big]
\end{displaymath}
where $G_{0}(\vec{p}) = \frac{-i}{p_{\mu} \gamma^{\mu}}$ and the vertices are
the $4 \times 4$ unit matrices denoted $\openone$.  Decoupling the denominator
in the usual way via Feynman parameters  we obtain
\begin{equation}\label{Meanfield}
\langle S^{+}_{s}(\vec{q}) S^{-}_{s}(\vec{q}) \rangle_{0} = -
\frac{\sqrt{q_0^2 + \mathbf{q}^2}}{16}
\end{equation}

\begin{figure}[tb]
\begin{center}
\includegraphics[width=65mm]{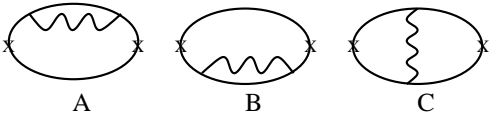}
\caption{Non-zero leading $1/N$ corrections to the staggered spin
correlation function. The {\bf x} denotes the vertex which is the $4 \times
4$ unit matrix in  the case of interest.}
\label{Pol1N}
\end{center}
\end{figure}

Note that $\mathbf{q}$ is measured with respect to $(\pi,\pi)$ in units of the
lattice spacing.  From expression (\ref{Meanfield}), we see that
the antiferromagnetic
correlations have been largely lost. The reason for this can be traced back to
the spin singlet formation in the mean-field sF (U1Cn01n) phase.
This motivates the inclusion of gauge fluctuations.  At order $1/N$ we have
the three non-vanishing diagrams depicted in Fig.[\ref{Pol1N}].  In order to
calculate the contribution of these diagrams we note that unlike the single
spinon spectral function the density density correlation is gauge invariant
and hence we can choose to work in the Landau gauge where the gauge propagator
reads
\begin{equation}\label{Landau_gauge}
D_{\mu \nu}(\vec{q}) = \frac{8}{N \sqrt{\vec{q}^2}}\big ( \delta_{\mu\nu} -
\frac{q_{\mu}q_{\nu}}{\vec{q}^2} \big )
\end{equation}
and we have the following expressions for the diagrams shown
\begin{widetext}
\begin{equation}
[\ref{Pol1N}(A)] + [\ref{Pol1N}(B)] = -\frac{1}{2}\int \frac{d^d k}{(2\pi)^d} 
\int
\frac{d^d q}{(2\pi)^d} \Tr \left [\openone
\frac{1}{i(p+k)_{\epsilon}\gamma^{\epsilon}} \openone
\frac{1}{ik_{\delta}\gamma^{\delta}}i\gamma^\mu
D_{\mu\nu}(\vec{q})\frac{1}{i(k+q)_\beta \gamma^\beta}i\gamma^\nu
\frac{1}{ik_\alpha \gamma^\alpha} \right ]
\end{equation}
\begin{equation}
[\ref{Pol1N}(C)] = -\frac{1}{4}\int \frac{d^d k}{(2\pi)^d} 
\int \frac{d^d q}{(2\pi)^d}
\Tr \left [\openone \frac{1}{i(p+k)_{\epsilon}\gamma^{\epsilon}}i\gamma^{\mu}
\frac{1}{i(p+k+q)_{\delta}\gamma^{\delta}}\openone
\frac{1}{i(k+q)_{\beta}\gamma^{\beta}}i\gamma^{\nu}
\frac{1}{ik_{\alpha}\gamma^{\alpha}}D_{\mu\nu}(\vec{q}) \right] 
\end{equation}

After performing the trace and integrating over $k$ in $d=3$, where the
integrals are convergent (see Appendix \ref{appB}), we arrive at
\begin{equation}
[\ref{Pol1N}(A)]+ [\ref{Pol1N}(B)] =
\frac{1}{N}\int \frac{d^d q}{(2\pi)^d} \bigg [ \frac{2\vec{p}\cdot\vec{q} +
\vec{p}^2}{|\vec{q}|^3 |\vec{p} + \vec{q}|} -
\frac{\vec{p}\cdot\vec{q}}{|\vec{p}|\vec{q}^2|\vec{p}+\vec{q}|} -
\frac{|\vec{p}|}{|\vec{q}|^3}\bigg]
\end{equation}
\begin{equation}\label{Pol1NC}
[\ref{Pol1N}(C)] = \\
 \frac{1}{N}\int \frac{d^d q}{(2\pi)^d}\bigg [\frac{2}{\vec{q}^2} +
\frac{|\vec{p}|}{|\vec{q}|^3} - \frac{\vec{p}^2 +
2\vec{p}\cdot\vec{q}}{|\vec{q}|^3|\vec{p}+\vec{q}|} -\frac{4
|\vec{p}|}{\vec{q}^2|\vec{p}+\vec{q}|} - \frac{\vec{q}^2\vec{p}^2 + 2
\vec{p}^4}{\vec{p}\cdot\vec{q}\vec{q}^2|\vec{p}||\vec{p}+\vec{q}|} \bigg ]
\end{equation}
\end{widetext}
In order to proceed we need to regularize the above integrals.  Because of the
dot-product appearing in the denominator of the last term in (\ref{Pol1NC}) it
is hard to use dimensional regularization. Hence we have set $d=3$ and
introduced an upper momentum cutoff $\Lambda$.  Thus performing the final
integrals over $\vec{q}$ we can extract the $log$-divergent term in the form
\begin{displaymath}\label{result}
[\ref{Pol1N}(A)]+ [\ref{Pol1N}(B)]+[\ref{Pol1N}(C)] = -\frac{8}{12 \pi^2
N}|\vec{p}| ln\bigg(\frac{\Lambda^2}{\vec{p}^2}\bigg)
\end{displaymath}

After combining this with the mean-field result (\ref{Meanfield}) and
continuing to real frequencies (see Appendix \ref{appB}) we obtain
\begin{widetext}
\begin{eqnarray}\label{sspin}
\Im\chi(\omega,\mathbf{q}) &\equiv& \chi^{''}(\omega,\mathbf{q}) \equiv\Im\langle S^{+}_{s}(\omega,\mathbf{q})
S^{-}_{s}(-\omega,-\mathbf{q}) \rangle \quad \quad \quad \quad \nonumber \\ &=&
C_{s}\frac{1}{2}sin(2\nu\pi)\Gamma(2\nu-2)
\Theta(\omega^2-\mathbf{q}^2)\left(\omega^2-\mathbf{q}^2\right)^{1/2-\nu} \\
 \nu &=& \frac{32}{3\pi^2 N} \nonumber
\end{eqnarray}
\end{widetext}
where $C_{s}$ 
is a constant depending on the physics at the lattice scale.  In the limmit $N
\rightarrow \infty$ this reduces to the mean-field result (\ref{Meanfield})

\begin{figure}[b]
\begin{center}
\includegraphics[width=70mm]{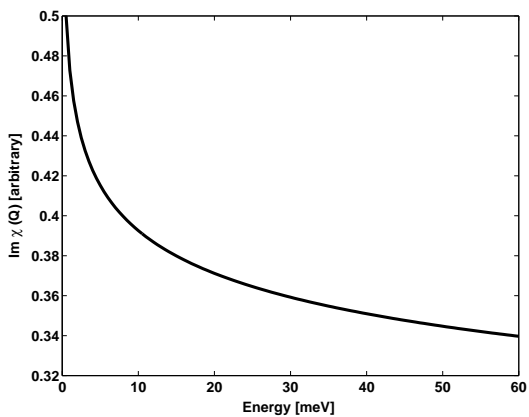}
\caption{Imaginary part of the spin susceptibility at $\mathbf{Q_{AF}}$. 
Note the divergence at small $\omega$}.
\label{chi_Q_ASL}
\end{center}
\end{figure}

{}From (\ref{sspin}) it is clear that the gauge fluctuations have reduced the
mean-field exponent. If we boldly set $N=2$ which is the physically relevant
case we find $\nu > 1/2$ signaling an antiferromagnetic instability.  
This is consistent with an earlier numerical result obtained in Ref. 
\onlinecite{Ivanov}. In
Fig.[\ref{chi_Q_ASL}] we plot the imaginary part of the spin susceptibility at
$\mathbf{q} = 0 \equiv \mathbf{Q_{AF}} = (\pi,\pi)$

This result is quite natural in the light of what has been said so far. The
gauge fluctuations arise from the constraint of no double occupancy. The
dynamics for the gauge field - in the very low doping limit assumed here - is
solely due to virtual spinons. It is the low energy spinon-antispinon pairs
having a nesting condition for scattering between the nodes (separated by
$\mathbf{Q_{AF}}$) in the sF (U1Cn01n) phase mean-field spectrum which are
responsible for the antiferromagnetic enhancement in the ASL phase.  Let us
emphasize again that this enhancement of spin correlations at $\mathbf{Q_{AF}} =(\pi,\pi)$ over the uniform component is protected, just like the gapless
$U(1)$ gauge fluctuations and the gapless spinons, by the U1Cn01n quantum
order

We have thus established algebraic behavior for the staggered spin
correlations. It is such an algebraic correlation that leads to the name ASL
that we assigned to the phase.

\subsection{Correlations near $(\pi,0)$} 

For this momentum transfer the spinons get scattered between the two different types of
nodes in the sF spectrum which yields the following expression for the
corresponding correlations 
\begin{eqnarray}
\langle S^{+}_{(\pi,0)}(\vec{q}) S^{-}_{(\pi,0)}(-\vec{q}) \rangle = \quad\quad\quad\quad\quad \\
 -\frac{1}{4}\int\frac{d\vec{p}}{(2\pi)^3} \Tr\big[\pmatrix{0&\sigma_{1}\cr\sigma_{1}&0\cr} G(\vec{p}) \pmatrix{0&\sigma_{1}\cr\sigma_{1}&0\cr}
G(\vec{p}-\vec{q}) \big] \nonumber
\end{eqnarray}

Having worked hard to obtain the anomalous dimension of the staggered spin
operator by brute force we will here resort to standard field theory
renormalization apparatus which allows for a more economical derivation of the
anomalous dimensions of composite operators.  In order to obtain the anomalous
dimension of the spin operator which is a fermion bilinear we need to obtain
its wavefunction renormalization in the form
\begin{equation}\label{ZS}
Z_{S} = Z_{\Psi}Z_{\Gamma}
\end{equation}
where $\Gamma$ is the relevant vertex and as we have seen above depends on the
momentum transfer ($\Gamma = \gamma_{0}, \openone, \pmatrix{0&\sigma_{1}\cr
\sigma_{1}&0\cr}$ near $(0,0), (\pi,\pi)$ and $ (\pi,0)$ respectively).

The spinon wavefunction renormalization is obtained from the self energy in
the usual way and evaluates in the Landau gauge to 
\begin{equation}\label{wavefunction_psi}
Z_{\Psi} = 1+\frac{4}{N}\frac{1}{3\pi^2}
\frac{\Gamma(\frac{3-d}{2})}{(M^2)^{(3-d)/2}}
\end{equation}
where $M$ is the renormalization scale and we have used dimensional regularization.

\begin{figure}[tb]
\begin{center}
\includegraphics[width=20mm]{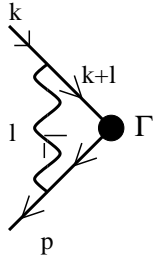}
\caption{Fermion-Fermion vertex}
\label{vertex}
\end{center}
\end{figure}

To obtain the vertex renormalization we need to evaluate the divergent part of
the 1-loop diagram depicted in Fig.[\ref{vertex}] which reads
\begin{displaymath}
\int\frac{d^3l}{(2\pi)^3}\frac{i\gamma^{\mu}(-i\gamma^{\epsilon})
(k+l)_{\epsilon}\Gamma (-i\gamma^{\delta})(l+p)_{\delta}i\gamma^{\nu} 
D_{\mu\nu}(l) }{(k+l)^2 (p+l)^2}
\end{displaymath}
where $\Gamma$ is the relevant vertex and $D_{\mu\nu}$ is given by
(\ref{Landau_gauge}).  Since the vertex correction diverges only
logarithmically we can simply evaluate the divergent part by setting all the
incoming momenta equal to $0$.  Hence the above simplifies to
\begin{eqnarray*}
\frac{8}{N}\int\frac{d^dl}{(2\pi)^d} 
\quad\quad\quad\quad\quad\quad\quad\quad\quad \\
\left[\frac{\gamma^\mu l_{\epsilon}\gamma^{\epsilon} 
\Gamma_{(\pi,0)}l_{\delta}\gamma^{\delta}\gamma^{\nu}\delta_{\mu\nu}}{l^5}  
- \frac{\gamma^\mu l_{\epsilon}\gamma^{\epsilon} \Gamma_{(\pi,0)}l_{\delta}
\gamma^{\delta}\gamma^{\nu}l_{\mu}l_{\nu}}{l^7}\right]
\end{eqnarray*}
where we now concentrate on the $(\pi,0)$ vertex.

This can be easily evaluated with the help of 
\begin{eqnarray*}
\gamma_{\epsilon}\Gamma_{(\pi,0)} = 
(1-2\delta_{\epsilon 2})\Gamma_{(\pi,0)}\gamma_{\epsilon} \\
\int\frac{d^dl}{(2\pi)^d}\frac{1}{l^3}|_{div} = 
\frac{1}{4\pi^2}\frac{2}{3-d}
\end{eqnarray*}
and results in
\begin{equation}
Z_{\Gamma_{(\pi,0)}} = 1 - \frac{4}{N}\frac{1}{3\pi^2}
\frac{\Gamma(\frac{3-d}{2})}{(M^2)^{(3-d)/2}}
\end{equation}
which together with equations (\ref{ZS}) and (\ref{wavefunction_psi}) gives
\begin{equation}
Z_{S_{(\pi,0)}} = 1
\end{equation}
Hence the spin operator at $(\pi,0)$ doesn't pick up any anomalous dimension
from the gauge field interactions.  It is not hard to check within the above
outlined calculation that for the other two vertices corresponding to the
uniform and staggered spin operator, we recover the results discussed in the
previous sections which gives a nice check on the brute force calculation.

\section{Perturbative stability of the $\text{sF}$ ($\text{U1Cn01n}$) phase}
\label{ASLPSG}

An important technical detail was hidden in the above calculation. In this
section, we will expose this issue.

The sF phase contains 2 families of 4-component massless Dirac fermions, which
couple to massless $U(1)$ gauge fluctuations. At the perturbative level, the
interaction can generate many possible counter terms (in this case the
self-energy terms) which can \\
(A) generate a mass for the Dirac fermions,\\
(B) generate a finite chemical potential for the Dirac fermions and change the Fermi points into Fermi pockets,\\
(C) shift the crystal momenta of the Dirac fermions from $(\pm \pi/2, \pm
\pi/2)$ to some other values.\\
In our perturbative calculation in the last section, we have assumed that non
of the above counter terms are generated by the interaction.

In a generic interacting theory, we used to believe that all the counter terms
that are consistent with symmetries will be generated. All of the above three
types of counter terms are consistent with the underlying lattice symmetries and are hence allowed by the symmetries. In fact they do appear in our calculation of
the continuum theory as cut-off dependent terms. It appears that the results in
 the last section are incorrect since the important counter terms are ignored.
Those counter terms, if present, will drastically alter our previous results.

As was stressed in this paper and in Ref.\onlinecite{Wqos} and Ref.\onlinecite{Wqo}, the
sF phase, as a quantum spin liquid, is not only characterized by its symmetry
but also by its quantum order. The quantum order in the sF phase is
characterized by a particular PSG -- U1Cn01n. The mean-field ansatz of the sF
phase is invariant under the transformations of the U1Cn01n PSG.  As pointed
out in Ref.\onlinecite{Wqos} and Ref. \onlinecite{Wqo}, the perturbative fluctuations
around the mean-field ground state may deform the mean-field ansatz. Those
deformations correspond to the counter terms. However, in our case, it is
incorrect to use symmetries to determine which counter terms are allowed. One
should use the PSG to determine the allowed counter terms. This is because
perturbative fluctuations can only deform the ansatz in such a way that the
deformed ansatz remains invariant under the same PSG.  It was shown\cite{Wqo}
that the above three counter terms are forbidden by the U1Cn01n PSG. This is
why we can drop them in our calculation.

When we calculate the effects of interactions in a continuum theory, we introduce
a high energy/short distance cut-off. This cut-off destroys the 
structure of the underlying quantum order. Thus it is not surprising that the counter terms
forbidden by the quantum order show up in continuum theory as cut-off dependent
terms. To restore the quantum order that existed in the underlying lattice
theory, we can simply drop all the forbidden counter terms in our calculation
within the continuum theory. We see that the PSG and the concept of quantum order play an important role even in calculations within continuum theories. It is the understanding of the
PSG and quantum order that makes sensible calculations in the continuum limit
possible. In fact, the theory of quantum order is partly motivated by the
above issue of the counter terms.

In summary, the perturbative fluctuations around the mean-field sF (U1Cn01n)
state cannot change the quantum order and cannot generate energy gaps for the
$U(1)$ gauge field and the spinons\cite{Wqo}.  Thus the gapless $U(1)$ gauge
fluctuations and the gapless spinons are protected by the U1Cn01n quantum
order present in the sF (U1Cn01n) phase. They remain gapless even when the
interaction is finite all the way down to zero energy. The interacting gapless
excitations lead to many unusual properties of underdoped cuprates, such as
the non-Fermi liquid behavior of the normal metallic state, the broad
electron spectral function\cite{ASL}, and the diverging AF spin fluctuations
in the presence of the pseudo-gap.  The U1Cn01n quantum order in the sF phase
not only protects the gapless excitations, it also protects the momentum of
the gapless excitations.  For example the spin-1 gapless excitations can only
appear near $\mathbf{k} =(0,0)$, $(\pi, \pi)$, $(0, \pi)$, and $(\pi, 0)$.

We would like to point out that although the ASL can be a stable quantum phase
in the large $N$ limit,\cite{IoffeLarkin,Wqo} for the real $N=2$ case,
non-perturbative instanton effects cause an instability.  Thus at low energies
the ASL will change into some other state, such as the $d$-wave
superconducting state, the AF state, stripe states, or even a $Z_2$ spin
liquid state.  Since the U1Cn01n quantum order in the sF phase requires the
spin-1 gapless excitations to appear at $\mathbf{k} =(0,0), (\pi, \pi),
(\pi,0), (0,\pi)$, the shift of the low frequency neutron scattering peak
observed in experiments
\cite{YMK8886,BEH8968,CAM9191,FKM9713,BFR9739,YLK9865,FBS0073}
indicates a transition from the ASL to
some other state at low temperatures.  Studying how the momentum of spin-1
gapless excitations shifts away from $(\pi,\pi)$, $(\pi,0)$ and $(0,\pi)$ will
allow us to experimentally identify the low temperature phase.  In the last
section, we have studied the spin fluctuations in the ASL.  In the following section, we
will study the spin fluctuations in the low temperature phase. 


\section{The fate of the ASL}\label{fate}

In section \ref{susceptibility}, we showed how the ASL physics reconciles the
pseudo-gap formation ({\it ie} the condensation into spin singlets within the
RVB picture) with enhanced dynamical antiferromagnetic fluctuations. At lower
energies the ASL is unstable - and an important question is how the ASL
evolves in the underdoped regime into the superconducting state as we reduce
temperature. 

In the following we are going to address this question via first analyzing the
effect of the opening of a gap in the gauge fluctuations on the staggered spin
correlations. Intuitively we should expect that for energies below the mass
gap the spin correlations should be given by mean-field correlations - which is
indeed the case. Having established the profound effects of this gap formation
we take a more careful look at the change of the mean-field correlations on
going from the sF (U1Cn01n) phase into the fermion pairing state which will allow us to
address the question of incommensurate spin fluctuations seen at low
frequencies and temperatures in the cuprates.  Let us now proceed to consider
the effect of giving a mass to the gauge fluctuations.
In the context of the spin correlations considered here, we would thus expect
the destruction of the antiferromagnetic enhancement below the mass gap and a
resurfacing of the singlet character of the correlations in the mean-field
sF (U1Cn01n) phase.

To implement the mass gap formation we take the following phenomenological
form for the gauge propagator\cite{ASL}
\begin{displaymath}
D_{\mu \nu}(\vec{q}) = \frac{8}{N \sqrt{\vec{q}^2 + m^2}}\big (
\delta_{\mu\nu} - \frac{q_{\mu}q_{\nu}}{\vec{q}^2} \big )
\end{displaymath}
The calculation then goes through as above (with one subtlety in the analytic
continuation discussed in the Appendix \ref{appB}) resulting in the following
expression for the staggered correlation

\begin{eqnarray}
\label{resultm}
&& \Im\langle S^{+}_{s}(\omega,\mathbf{q}) S^{-}_{s}(\omega,\mathbf{q}) \rangle
\\
&=&
\Theta(\omega^2 -\mathbf{q}^2)\Theta\left[m^2 +\mathbf{q}^2 -\omega^2)\right] 
C_m \frac{\sqrt{\omega^2-\mathbf{q}^2}}{m^{2\nu}} 
\nonumber  \\
&+&
\Theta\left[ \omega^2 - \mathbf{q}^2-m^2\right]
C_m\frac{\sqrt{\omega^2-\mathbf{q}^2}}{\left(\sqrt{\omega^2-\mathbf{q}^2} 
+ \sqrt{\omega^2 -\mathbf{q}^2-m^2}\right)^{2\nu}} 
\nonumber 
\end{eqnarray}
where
\begin{displaymath}
\nu = \frac{32}{3\pi^2 N} 
\end{displaymath}

\begin{figure}[tb]
\begin{center}
\includegraphics[width=70mm]{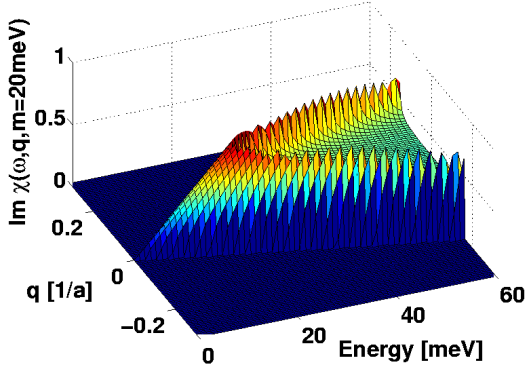}
\caption{Scan of the imaginary part of the spin susceptibility in arbitrary
units for $\mathbf{q} = -\pi/10 \ldots \pi/10$ and $\omega = 0 \ldots
60meV$. We have chosen $m = 20meV$ in this plot }
\label{chi_oq_scan}
\end{center}
\end{figure}

\begin{figure}[tb]
\begin{center}
\includegraphics[width=70mm]{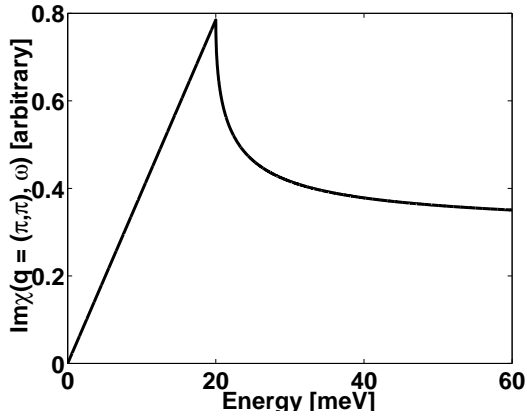}
\caption{Imaginary part of the spin susceptibility at 
$\mathbf{Q_{AF}}$ where $m = 20 meV$ 
notice the recovery of the linear mean-field result below the
massgap}
\label{chi_Q_m}
\end{center}
\end{figure}

In Fig.[\ref{chi_oq_scan}] 
we show the resulting spectra with $m = 20meV$. Fig.[\ref{chi_Q_m}] depicts the spectrum at
the antiferromagnetic ordering wavevector $\mathbf{Q_{AF}}$. As expected the
opening of the massgap in the gauge fluctuations has restored the mean-field
result and consequently suppressed the antiferromagnetic enhancement seen in
Fig.[\ref{chi_Q_ASL}].

{}From this analyzes we see that the spin correlations at low frequencies are
described by the underlying mean-field . In the above calculation we have
assumed that the only effect of the destruction of the massless $U(1)$ gauge
structure in the sF phase is the formation
of the $U(1)$ mass gap which leaves the underlying mean-field correlations in 
tact at low energies. On a
more profound level however the massless nature of the $U(1)$ gauge structure
in the sF state is a manifestation of the U1Cn01n quantum
order\cite{Wqo} present in this state (see introduction and section
\ref{ASLPSG}). Thus it is not possible to simply break the gauge structure
without affecting the underlying mean-field ansatz as we have assumed above.
If we reverse the above logic, this implies that the low frequency spin
correlations which reflect the underlying mean-field ansatz can be used as an
experimental probe into the corresponding quantum order/gauge structure
accompanying the mean-field correlations. 

The ASL described by the sF (U1Cn01n) state may change into several different
states at low temperatures. The low energy spin correlation can have different
behaviors in those low temperature states. In the following, we will
discuss one particular low temperature state to gain some intuition on the
behavior of the low energy spin correlation in the low temperature phases.

Within the $SU(2)$ slave boson theory,
the mean-field phase which is energetically favored over the sF (U1C01n)
phase at low temperatures is the so called d-wave pairing state
\cite{fourauthor}. In order to keep this section reasonably self contained we
would like to outline some of the steps leading to the mean-field correlations
discussed in the following.  As was mentioned above within the SU(2) slave
boson formulation the physical electron operator is represented as:
\begin{eqnarray}\label{SU2rep}
c_{\uparrow i}=\frac{1}{\sqrt{2}}h^{\dag}_{i}\psi_{\uparrow i} =
\frac{1}{\sqrt{2}}\big(b^{\dag}_{1i}f_{\uparrow
i}+b^{\dag}_{2i}f^{\dag}_{\downarrow i}\big) \\ \nonumber
c_{\downarrow i}=\frac{1}{\sqrt{2}}h^{\dag}_{i}\psi_{\downarrow i} =
\frac{1}{\sqrt{2}}\big(b^{\dag}_{1i}f_{\downarrow
i}-b^{\dag}_{2i}f^{\dag}_{\uparrow i}\big) \\ \nonumber
\end{eqnarray}
where the following SU(2) doublets were introduced
\begin{eqnarray}
\psi_{\uparrow i} = {f_{\uparrow i} \choose f^{\dag}_{\downarrow i}}, \quad
\psi_{\downarrow i} = {f_{\downarrow i} \choose -f^{\dag}_{\uparrow i}},
\quad
h_{i} = {b_{1i} \choose b_{2i}} \nonumber
\end{eqnarray}

The $\psi_{\uparrow i}, \psi_{\downarrow i}$ are the two fermion fields
representing the destruction of a spin up and spin down on site i respectively
and $h_{i}$ is the doublet of bosonic fields keeping track of the doped holes.
Putting this representation into the t-J Hamiltonian
\begin{displaymath}
H = \sum_{(ij)}\big[ J(\vec{S}_{i}\cdot\vec{S}_{j} - \frac{1}{4}n_{i}n_{j})
- t(c^{\dag}_{\sigma i}c_{\sigma j} + h.c.)\big]
\end{displaymath}
yields on performing a Hubbard-Stratonovich transformation to the appropriate
bosonic bond variables the following partition function \cite{fourauthor}
\begin{eqnarray}
Z &=& \int DhDh^{\dag}D\psi^{\dag} D\psi
D\vec{a}_{0}DUe^{-\int_{0}^{\beta}L} \\
L &=& \frac{\tilde{J}}{2}\sum_{<ij>}Tr[U^{\dag}_{ij}U_{ij}] +
\frac{1}{2}\sum_{i,j,\sigma}\psi^{\dag}_{\sigma
i}(\partial_{\tau}\delta_{ij}+\tilde{J}U_{ij})\psi_{\sigma j} \nonumber \\
 &+& \sum_{il}a^{l}_{0i}\big(\frac{1}{2}\psi^{\dag}_{\sigma
i}\tau^{l}\psi_{\sigma i} + h^{\dag}_{i}\tau^{l}h_{i}\big) \nonumber \\
 &+&
\sum_{ij}h^{\dag}_{i}\big((\partial_{\tau}-\mu)
\delta_{ij}+\tilde{t}U_{ij}\big)h_{j}
\end{eqnarray}

The $\vec{a}_{0}$ fluctuations incorporate the projection to the space of
SU(2) singlets within the above representation for the electron operators
(\ref{SU2rep}). Furthermore note that\cite{UbbensLee} $\tilde{J}=3J/8$, 
$\tilde{t}=t/2$ and the matrix $U_{ij}$ in the form
\begin{displaymath}
U_{ij} = \pmatrix{-\chi_{ij}^{*}&\Delta_{ij} \cr
\Delta^{*}_{ij}&\chi_{ij}\cr}
\end{displaymath}
contains the Hubbard-Stratonovich fields which classify the part of the phase
diagram we are looking at. The mean-field phase diagram is found by minimizing
the free energy for a given number of particles with respect to the bond
variables $U_{ij}$.

The d-wave pairing state can be represented as
\begin{eqnarray}\label{SCstate}
U_{i,i+\hat{x}} &=& -\tau^{3}\chi + \tau^{1}\Delta \nonumber \\
U_{i,i+\hat{y}} &=& -\tau^{3}\chi - \tau^{1}\Delta \nonumber \\
a_{0}^{3} &\neq& 0 
\end{eqnarray} 

The key difference to the sF (U1Cn01n) phase is the appearance of a finite
$a_{0}^3$ which acts as a chemical potential for the spinons. Without this
term the above ansatz is $SU(2)$-gauge equivalent to the sF (U1Cn01n)
ansatz\cite{fourauthor}. It is shown in Ref.\onlinecite{Wqo} that the appearance of such
a chemical potential term is not consistent with the quantum order of the sF
(U1Cn01n) phase and hence signals the presence of a different quantum order
accompanying the d-wave pairing state.  It is precisely a nonzero $a_{0}^3$
which gives rise to an anomalous fermion fermion pairing which when combined
with holon condensation leads to the d-wave superconducting state. As we shall
see in the following, $a_{0}^3$ is also responsible for the incommensurate
spin response at the lowest frequencies shifting the peak away from
$\mathbf{Q_{AF}} = (\pi,\pi)$.

Taking the above ansatz equation (\ref{SCstate}) we can calculate the spin-spin
correlations at mean-field level (see Appendix \ref{appE}) to obtain for the
imaginary part of the spin correlation near $\mathbf{Q_{AF}} = (\pi,\pi)$
\newpage
\begin{widetext}
\begin{eqnarray}
&& \Im\langle S^{+}(\omega,\mathbf{Q_{AF} + k}) S^{-}(-\omega,-\mathbf{Q_{AF}-k}) \rangle \nonumber
\\
&=& \frac{1}{64 v_{f}v_{2}} \left
\{\theta\left[ \omega^2 - (v_f k_1 +2 a_0^3)^2 - v_2^2k_2^2 \right]  
\sqrt{\omega^2 - (v_f k_1 +2 a_0^3)^2 - v_2^2k_2^2} +\right.
\nonumber \\
&& \quad \quad \quad  \theta\left[ 
\omega^2 - (v_f k_1 - 2 a_0^3)^2 - v_2^2k_2^2 \right]  
\sqrt{\omega^2 - (v_f k_1 - 2 a_0^3)^2 - v_2^2k_2^2} + 
\nonumber \\
&& \quad \quad \quad \theta\left[ \omega^2 - (v_f k_2 +2 a_0^3)^2 - v_2^2k_1^2 
\right]  
\sqrt{\omega^2 - (v_f k_2 +2 a_0^3)^2 - v_2^2k_1^2} + 
\nonumber \\
 && \left. \quad \quad \quad \theta\left[ 
\omega^2 - (v_f k_2 - 2 a_0^3)^2 - v_2^2k_1^2 \right]  
\sqrt{\omega^2 - (v_f k_2 - 2 a_0^3)^2 - v_2^2k_1^2}\right\} 
\\
k_1 &\equiv& \frac{k_x+k_y}{\sqrt{2}} \quad
k_2 \equiv \frac{-k_x + k_y}{\sqrt{2}}\quad
v_{f} \equiv 2\sqrt{2}aJ\chi \quad
v_{2} \equiv 2\sqrt{2}aJ\Delta \nonumber
\end{eqnarray}
\end{widetext}

\begin{figure}[tb]
\begin{center}
\includegraphics[width=70mm]{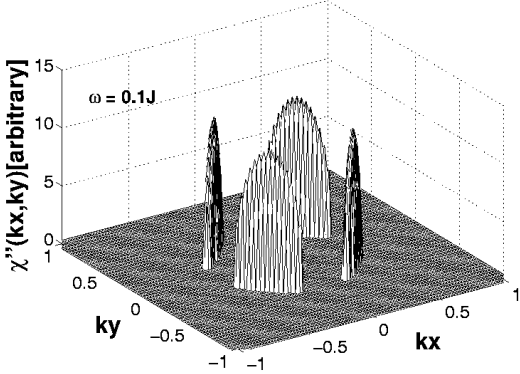}
\caption{Wave vector scan of $\chi''(k_x,k_y, \omega = 0.1J)$. Notice the
incommensurate pattern at low energy with amplitude peaks shifted along the
diagonals.\cite{BrinckmannLee} $(0,0)$ corresponds to $\mathbf{Q_{AF}}$ and
$k_x,k_y$ are measured in units of $1/a$ with $a$ the lattice spacing} 
\label{SCchi12mev}
\end{center}
\end{figure} 

\begin{figure}[tb]
\begin{center}
\includegraphics[width=70mm]{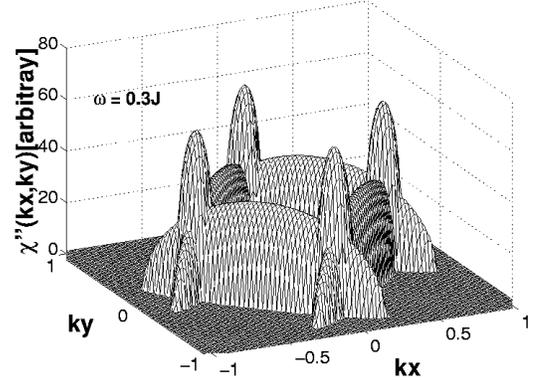}
\caption{Wave vector scan of $\chi''(k_x,k_y, \omega = 0.1J)$, $k_x,k_y$ are measured in units of $1/a$.
With $a_0^3$chosen to give $\delta a$ = 0.1 r.l.u. (2$\pi$ = 1 r.l.u.)\cite{Mook, Arai}
the value of $\omega_c$ is set by $\omega_c = \frac{4}{\sqrt{3}}
\left(\frac{v_{2}}{v_f}\right)\delta a J$. To compare with experiments we have
chosen the values $\frac{v_2}{v_f} = 1/7$\cite{Krishana} which gives $\omega_c
= 0.2 J$. Notice how the overlapping ridges lead to the horizontal
incommensuration.} 
\label{SCchi36mev} 
\end{center} 
\end{figure} 

In Fig.[\ref{SCchi12mev}] we plot the resulting spectrum for a frequency 
$\omega<\omega_c\equiv\frac{4}{\sqrt{3}}\left(\frac{v_2}{v_{f}}\right) a_0^3$ 
where the intensity is peaked around four points shifted diagonally away from
$Q_{AF}$. In contrast to this in Fig.[\ref{SCchi36mev}] we depict a scan for
$\omega > \omega_c$ the spectrum is peaked at four points shifted horizontally
$(\pi\pm\delta,\pi), (\pi,\pi\pm\delta)$ where the ridges - growing out
perpendicular to the diagonals - overlap. $\delta$ is related to $a_0^3$ via
$\delta a = \frac{a_0^3}{J}$ with $a$ the lattice spacing.   It should be
noted that in the $\omega \rightarrow 0$ the peaks are always located along
the diagonal, as expected, for in this regime the response is dominated by
creation of particle hole pairs connecting the Fermi points. Interestingly for
intermediate energies the peak intensity in the spectrum is shifted
\emph{horizontally} away from $\mathbf{Q_{AF}}$ even without one dimensional
phenomenology.

\section{Application to high $T_c$}\label{Discussion}

Before launching into a discussion of how the above mentioned physics embodied
in the ASL might shed some light on the cuprate spin physics we need to
briefly digress and establish some nomenclature. As we shall see in the
following each experimental community has their own set of
pseudo/spin-pseudo-gaps and it is easy to get lost in the confusing
terminology. Here we adopt the following convention: 

\emph{pseudo-gap} phase denotes the part in the cuprate phase diagram where
angle resolved photoemission experiments observe a large gap in the single
particle spectrum at $(\pi,0)$

The term \emph{spin-pseudo-gap} will be used in conjunction with the spectrum
at the antiferromagnetic wavevector $\mathbf{Q_{AF}} = (\pi,\pi)$ to denote
the frequency range of reduced spectral weight below the peak above
$T_{c}$ (which on cooling below $T_{c}$ shifts to higher frequencies and
becomes the ``resonance'' - see discussion below).

Over the years NMR and inelastic neutron scattering (INS) have painted a very
interesting picture for the spin correlations in the cuprates starting with
the discovery of the by now famous $41meV$ ``resonance'' peak by J.
Rossat-Mignod {\it et al} \cite{RM} in the superconducting state of near
optimally doped YBCO. In addition to this resonance mode which appears
exclusively in the superconducting state there is  now evidence for precursory
piling up of spectral weight in the energy region of the ``resonance'' even in
the  normal state of underdoped cuprates.  The energy at which this
enhancement occurs in the normal state is sometimes referred to as the
``spin-pseudo-gap'' energy by INS workers \cite{Bourges9901333} (hence our
convention). From the perspective of spin fluctuations INS is clearly the most
powerful probe giving both dynamical and wavevector dependent information
about the spectrum.

 NMR on the other hand probes particular regions in wavevector space with
Knight shift probing the uniform susceptibility - $\propto \chi(\mathbf{q} =
0,\omega \sim 0)$ - and spin lattice relaxation rate probing $\sim \frac{\Im
\chi(\mathbf{q}_{C_u,O},\omega)}{\omega}|_{\omega \rightarrow 0}$ where
$\mathbf{q}_{C_u} = \mathbf{Q_{AF}}$ and $\mathbf{q}_{O} = \mathbf{0}$ are the
characteristic wavevectors determining the spin lattice relaxation at the
planar $C_u$ and $O$ sites respectively.  In addition to the above mentioned
``spin-pseudo-gap'', NMR has identified two characteristic temperatures. One
is the reduction of the Knight shift below a temperature which we will denote
by $T_{N}$ which indicates the loss of spectrum for $\chi(\mathbf{q} =
0,\omega \sim 0)$ below this characteristic temperature.  Furthermore there is
a second \emph{lower} scale which shows up in the spin lattice relaxation rate
for $C_u$ as a peak in $\frac{1}{T_{1}T}(T)$ which we shall call $T_{T_{1}}$.
Notice that $T_{T_{1}}<T_{N}$ but both of these temperatures are above the
superconducting transition temperature $T_{c}$ in underdoped samples.  From
the perspective of mean-field theory it is impossible to explain these
observations\cite{TKF}.  Within the mean-field picture the reduction of the
Knight shift below $T_{N}$ can be related to the reduction of spin fluctuations
on condensing into the spin singlet sector identified as the sF (U1Cn01n) phase above.
However there is then no way to explain the difference in the spin lattice
relaxation rates observed for planar $C_u$ and $O$.  Whereas
$\frac{1}{T_{1}T}(T)_{C_u}$ {\emph increases} with decreasing temperature
between $T_{N} > T >T_{T_{1}}$,  $\frac{1}{T_{1}T}(T)_{O}$ reduces
monotonically below $T_{N}$.

Let us now discuss these interesting phenomena from the point of view of the
ASL physics described in the previous part.  As has been suggested by Kim and
Lee\cite{KimLee} the enhancement of $C_u$ over $O$ relaxations is expected to
depend on the inclusion of the gauge fluctuations. The rational for this
suggestion arises from the fact that $C_u-NMR$ probes $\mathbf{q} \sim
\mathbf{Q_{AF}}$-fluctuations which are strongly affected by the gauge
fluctuations whereas $O-NMR$ derives its main contributions from $\mathbf{q}
\sim \mathbf{0}$ which is protected from gauge fluctuations by current
conservation.  We have shown above that this enhancement indeed takes place
with $\Im\chi_{\mathbf{Q_{AF}},\omega \rightarrow 0}$ (see
Fig.[\ref{chi_Q_ASL}]) being strongly increased over the mean-field behavior
which determines the spin fluctuations at $\mathbf{q} = 0$.  Given the
different $T$- dependence for the spin lattice relaxation rates for
temperatures $T_{N} > T >T_{T_{1}}$ it is thus natural to identify this regime
with the ASL physics.

\begin{figure}[tb]
\begin{center}
\includegraphics[width=70mm]{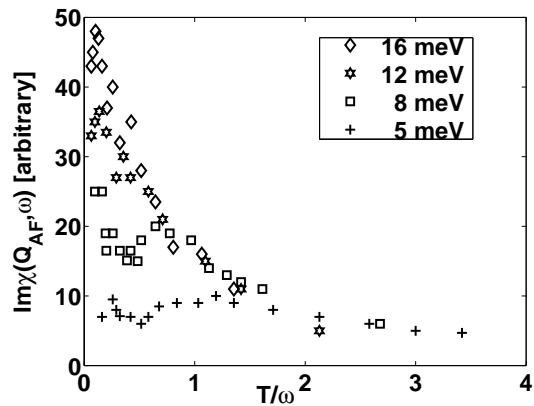}
\caption{Imaginary part of the spin susceptibility at $\mathbf{Q_{AF}}$ taken
from Sternlieb {\emph et al}. Note the temperature dependence of the
$\Im\chi$ for a fixed $\omega$ converges to a universal function of
$T/\omega$ at large $T$.}
\label{Sternlieb}
\end{center}
\end{figure}

Sternlieb {\emph et al}\cite{Sternlieb} have performed temperature dependent
neutron-scattering experiments on $YBa_{2}C_{u3}O_{6.6} (T_{c} = 53K)$ and
found a universal dependence of
\begin{equation}
\Im\chi(\omega) = \int \Im\chi(\mathbf{q},\omega)d\mathbf{q} \propto
\Im\chi(\mathbf{Q_{AF}},\omega)
\end{equation}
on the ratio $\omega/T$ above a temperature $T\sim 100K$. Their data (see
Fig.[\ref{Sternlieb}]) suggests that the temperature at which this universal
scaling appears increases with decreasing energy. In particular for $\omega =
5 meV$ the onset temperature is $T\sim 100K$. Below this characteristic
temperature (which again is well above $T_{c}$) their data shows a decrease in
spectrum which they associate with the opening of a spin-pseudo-gap below
$\omega_{g} \sim 10meV$. They also point out that their data and the fact that
the onset temperature of scaling increases with decreasing probe-frequency are
consistent with the anomalous behavior of $\frac{1}{T_{1}T}(T)_{C_u}$ which
probes $\omega \sim 0$ and increases down to $T=T_{T_{1}} \sim 150K$

\begin{figure}[tb]
\begin{center}
\includegraphics[width=70mm]{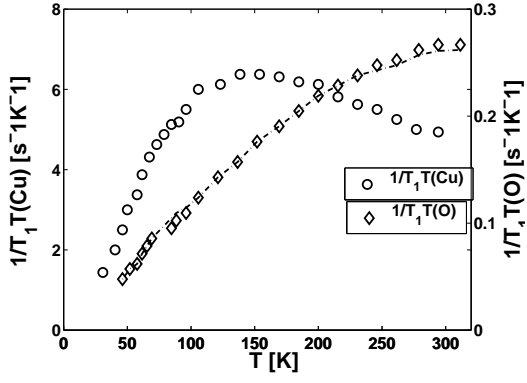}
\caption{Temperature dependence of $1/T_{1}T(C_u)$ and $1/T_{1}T(O)$ in
$YBa_{2}C_{u3}O_{6.63}$. The dotted-dashed line shows the temperature
dependence of the static susceptibility. The data is taken from
\cite{SlichterLAS} who quotes Takigawa's results}
\label{Slichter}
\end{center}
\end{figure}

As we argued above it is exactly the regime for $\omega > m$ at $T \sim 0$ or
$T>T_{m}$ at $\omega \sim 0$ that we associate with the ASL physics.  Under the
assumption of perfect scaling we can thus account for the difference in
$\frac{1}{T_{1}T}(T)$ between $C_u$ and $O$ quite naturally.  From
\begin{equation}
\frac{1}{T_{1}T} \propto \sum_{\mathbf{q}}
\frac{\Im\chi(\mathbf{q},\omega)}{\omega}|_{\omega \rightarrow 0}
\end{equation}
where for $C_u$: ($\sum_{\mathbf{q}} \sim \mathbf{Q_{AF}}$) due to the form
factor whereas $O$ probes the uniform spin fluctuations. In the ASL we can for
$C_u$ approximate
$\sum_{\mathbf{q}}\frac{\Im\chi(\mathbf{q},\omega)}{\omega}|_{\omega
\rightarrow 0} \sim \frac{\Im\chi(\mathbf{Q_{AF}},\omega)}{\omega}|_{\omega
\rightarrow 0}$ which via scaling results in
\begin{equation}
\frac{1}{T_{1}T}(C_u) \propto \frac{1}{T^{2\nu-1}} \quad 2\nu =
\frac{32}{3\pi^2}
\end{equation}
whereas $\frac{1}{T_{1}T}(T)_{O}$ falls monotonically as $T$ is reduced
following the uniform susceptibility (see Fig.[\ref{Slichter}]). As $2\nu -1
\sim 0$  this is consistent with the weak temperature dependence seen by
Sternlieb {\it et al} Fig.[\ref{Sternlieb}] at large $T/\omega$.

The ASL gets destroyed below the massgap  energy scale which we associate with
a temperature $T_{m}$.  Below this temperature scale both the $C_u$ and the $O$
spin lattice relaxation rates fall monotonically following the sF-singlet
correlations as $T$ is reduced. This is seen in experiments (see
Fig.[\ref{Slichter}])

Thus we associate the temperature  $T_{T_{1}}\sim 150 K$ where
$\frac{1}{T_{1}T}(T)_{C_u}$ shows the peak with $T_{m}$ the temperature below
which the massless $U(1)$ gauge structure (and thus the underlying quantum
order) becomes unstable and undergoes a ``transition'' (see discussion at the
end of this section)

\begin{figure}[tb]
\begin{center}
\includegraphics[width=70mm]{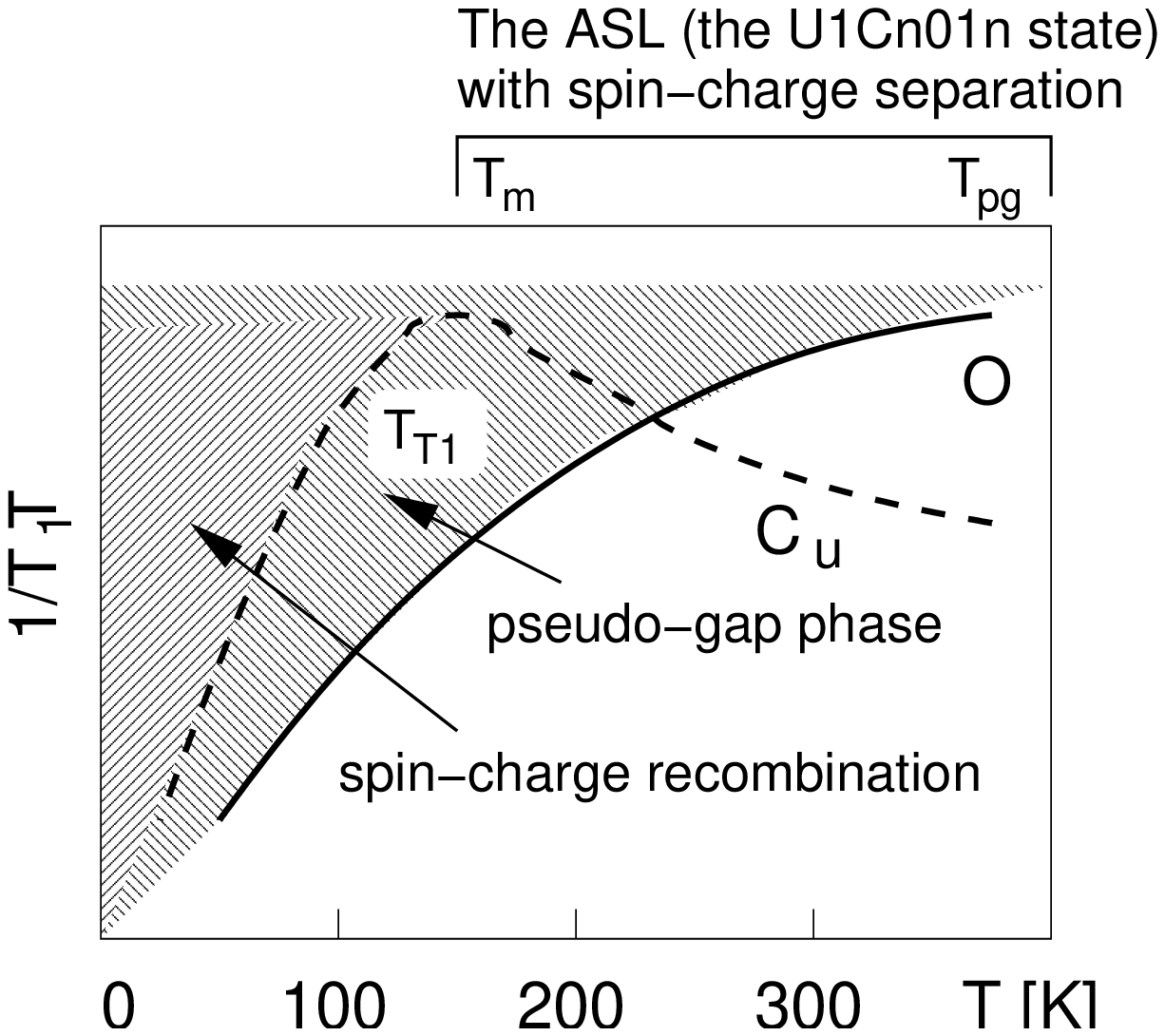}
\caption{Phase diagram implied by NMR experiments.}
\label{NMRphase}
\end{center}
\end{figure}

To summarize the above comparison between the ASL and NMR experiments, we
present the phase diagram Fig.[\ref{NMRphase}]. Below the pseudo-gap scale
$T_{pg}$,which we associate with $T_{N}$, the oxygen $1/T_1T$ starts to
decrease due to the opening of the pseudo-gap associated with the spin singlet
formation. However, the copper $1/T_1T$ keeps increasing as temperature
decreases, despite the small single particle density of states in the
pseudo-gap regime. This strange behavior can be explained very well by the ASL
due to the diverging AF spin fluctuations at low energies even in the presence
of the pseudo-gap. Below the temperature $T_m$, the $U(1)$ gauge field starts
to gain a gap and the enhancement in the AF spin fluctuations ceases to exist.
This causes the copper $1/T_1T$ to decrease with decreasing temperature
following the expected mean-field behavior. Thus the ASL described by the sF
state (or the U1Cn01n state) appears between $T_m$ (associated with
the experimental temperature scale $T_{T_{1}}$) and $T_{pg}$ (associated with
the experimental scale $T_{N}$). Below $T_m$, the ASL will change into another
state whose nature will be discussed later.

\begin{figure}[tb]
\begin{center}
\includegraphics[width=70mm]{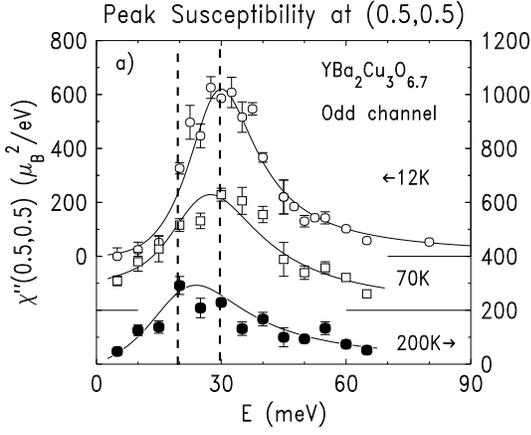}
\caption{Energy dependence of peak intensity at $\mathbf{Q_{AF}}$ as a function
of temperature taken from Fong {\it et al}\cite{FKM9713}. Note the shift
in the peak position on decreasing the temperature below $T_{c} = 67K$ (the
dashed lines were added by the current authors as a guide to the eye - see
main text). }
\label{Fong1}
\end{center}
\end{figure}

\begin{figure}[tb]
\begin{center}
\includegraphics[width=70mm]{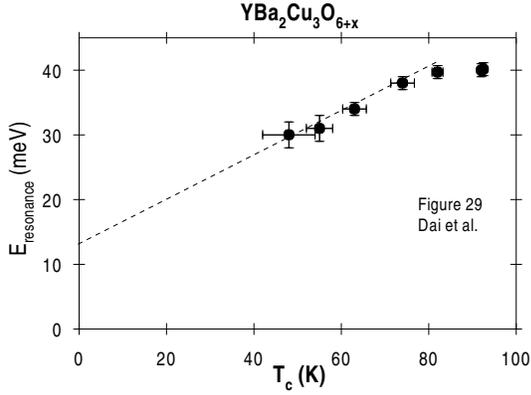}
\caption{The energy of the peak in the superconducting state - also called
the {\emph resonance} energy $E_{resonance}$ as a function of $T_{c}$ in
underdoped $YBCO$. Data taken from Dai {\it et al}\cite{Dai}. By
extrapolating the linear relationship down to $T_{c} = 0$ (dashed line added as
a guide) we extract an estimate for the mass scale related to
topological gauge fluctuations}
\label{Dai1}
\end{center}
\end{figure}

Let us now discuss the INS results which in their own right paint a very
interesting picture of the spin fluctuations in the cuprates. First we
consider the spectra at the antiferromagnetic ordering wavevector as a
function of temperature and energy. Already well above the superconducting
transition in the normal state, INS data show a marked increase in spectral
weight at a finite characteristic frequency $E_{spg}$ and a reduced spectral
weight for $\omega < E_{spg}$ ( the spin-pseudo-gap see Fig.[\ref{Fong1}]
bottom curve and compare Fig.[\ref{chi_Q_m}]). On cooling below $T_{c}$ this
gap increases in size and the corresponding peak shifts to higher energies as
depicted in Fig.[\ref{Fong1}] where we have added the two dashed lines to the
data taken by Fong {\it et al}\cite{FKM9713} to guide the readers eye.
{}From the perspective of the ASL liquid the ``normal state'' above $T_{c}$
corresponds to the phase where the gauge field is in the massive phase due to
topological fluctuations. From the data by Dai {\it et al}\cite{Dai} see
Fig.[\ref{Dai1}] we estimate the mass scale ({\it i.e.} the energy-gap scale)
due to instantons to be of the order of $10-15meV$ which ties in nicely with
the spin-pseudo-gap scale $\sim 10meV$ estimated by Sternlieb.

\begin{figure}
\begin{center}
\includegraphics[width=70mm]{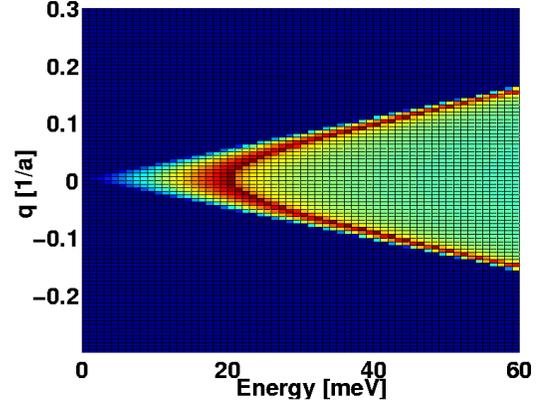}
\caption{
Contour plot of $\Im\chi$ - notice the ``spin-pseudo-gap'' at $\mathbf(q) =
0 \equiv \mathbf{Q_{AF}} = (\pi/a,\pi/a)$ for energies below $m = 20meV$ and the
dispersion above $m$.}
\label{chi_oq_scan_top}
\end{center}
\end{figure}

\begin{figure}[tb]
\begin{center}
\includegraphics[width=70mm]{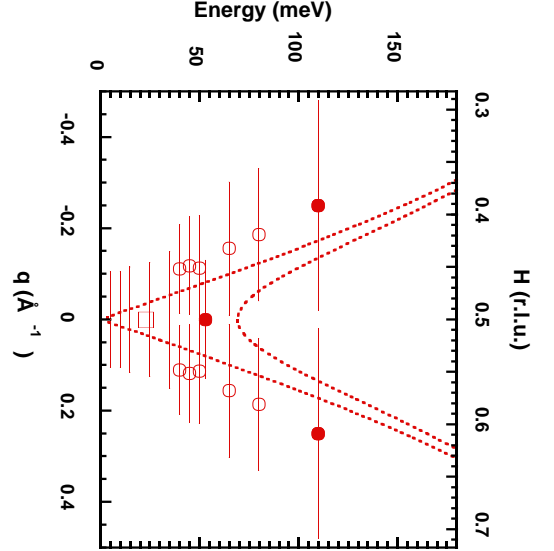}
\caption{Spin excitation spectrum for odd (open symbols) and even (closed
circles) excitations at 5K. The open square indicates the energy of the
maximum of the odd susceptibility. The dotted lines correspond to the
spin-wave dispersion relation in the insulating antiferromagnetic state with
$J_{||}$ = 120meV. This Fig is taken from Bourges {\it et al}
\cite{BFR9739}}
\label{Bourges2}
\end{center}
\end{figure}

The shift of this peak to higher energies on cooling below $T_{c}$ can be
qualitatively understood as the contribution to the mass scale by condensed
bosons via the Anderson-Higgs mechanism. This also explains the linear
relationship between $E_{resonance}$ (where $E_{resonance}$ denotes the energy
of the peak intensity in the superconducting state) and $T_{c}$ (both being
proportional to $x$ - the hole doping concentration) as observed by Dai{\it et
al} in Fig.[\ref{Dai1}] for underdoped $YBCO$. Furthermore we can account for
the very different doping dependence of the normal state maximum (determined
by the instanton scale with weak doping dependence) versus the doping
dependence of  $E_{resonance}$ (determined by the condensed bosons $\propto
x$) as was stressed in a recent article by Ph.  Bourges\cite{Bourges0009373}

{}From this understanding of the ASL, we can conclude immediately that, $T_m$
the temperature of the spin-pseudo-gap formation (and hence $T_{T_{1}}$)
should be rather insensitive to changes in doping. Contrast this with the
doping dependence of $T_{pg}$ the formation of the pseudo-gap in the single
particle spectrum which  as $x\to 0$ will get as large as the spin-wave band
width, since in that limit the ASL will be described by the $\pi$-flux phase.

Another interesting point to discuss about the INS data is their momentum
dependence. In Fig.[\ref{Bourges2}] we show the data by Bourges {\it et al}
taken on $YBa_{2}C_{u3}O_{6.5} (T_{c} = 52)$ which shows the spin excitation
spectrum for odd (acoustic) and even (optical) excitations at high energies.
The dotted lines correspond to the spin-wave dispersion relation in the
insulating antiferromagnetic state. The dispersive behavior compares nicely
with Fig.[\ref{chi_oq_scan_top}] the correlations in the ASL liquid above the
mass scale $m\sim 20 meV$


So far we have argued for the strange phenomenology of the pseudo-gap
phase to be tied to the physics embodied in the ASL. It is however clear that
the ASL physics has to give way to the superconducting state at low
temperatures. An important question is how this transition happens.  Let us
then briefly mention three plausible scenarios for how the ASL goes over to
the superconducting state. In other words, we would like to understand how the
$U(1)$ gauge field gains a mass term which leads to the destruction of the ASL
state.

In the first picture the mass of the $U(1)$ gauge field is generated via
confinement due to instantons.  After the opening of the energy gap, there is
no residual unbroken gauge structure left at low energies. Thus the
confinement can also be referred to as $U(1)$ gauge structure breaking down to
$Z_1$.  ($Z_1$ gauge structure means no gauge structure.) In the confinement
phase ({\it i.e.} after the $U(1)$ gauge field is gapped), the spinons and
holons recombine into electrons which appear as the relevant degrees of
freedom at low energies.  Thus we also call the first picture spin-charge
recombination.  This in particular would suggest the existence of well defined
quasiparticles at low energies (as probed by ARPES) even above the
superconducting transition temperature $T_c$. (this should also be accompanied
by $T^2$ resistance in d.c.  conductivity). Furthermore the binding of spinons and holons
will gives rise to Fermi arcs.\cite{WenLee} 

In the second picture, the $U(1)$ gauge structure also breaks down to the
$Z_1$ gauge structure, but now via holon condensation. In this case, $T_m=T_c$
and the ASL directly transforms into the $d$-wave superconducting state. There
are no well defined quasi particles above $T_c$.

A third scenario might be the breaking of the $U(1)$ gauge structure down to a
$Z_2$ gauge structure via the condensation of spinon bi-linear terms which do
not break any symmetries.\cite{WenZ2} (a $Z_2$ gauge structure was also found
in a slave-fermion description of spin systems via the condensation of boson
bi-linear terms which break $90^\circ$ rotation symmetry.\cite{RS} The $Z_2$
gauge structure can also be obtained via the condensation of bound states of
double vortices.\cite{BFN})  The breaking from $U(1)$ to $Z_2$ also results in
a mass for the $U(1)$ gauge field and leads to a new $Z_2$ spin liquid.  The
transition to this new topological/quantum order \cite{WenZ2,Wqo} implies
the appearance of {\em true} spin charge separation since the $Z_2$ gauge
interaction is only short ranged.\cite{WenZ2,RS,BFN} (Senthil experiment)  

\begin{figure}[tb]
\begin{center}
\includegraphics[width=80mm]{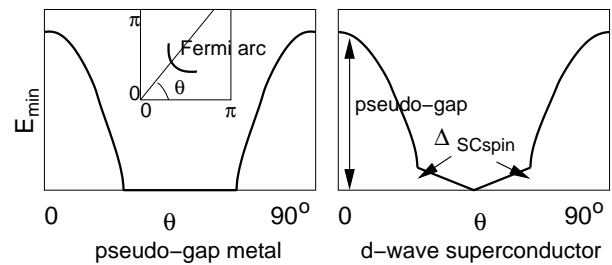}
\end{center}
\caption{Minimal quasiparticle energy $E_{min}$ along the line in the $\theta$
direction. The left panel depicts the pseudo-gap metallic state and shows a
finite Fermi arc. The right panel is for the superconducting state.
In the superconducting state, the quasiparticle may have four bands (instead of
the usual two for the BCS superconductor). For details, see \cite{WenLeeSC}.}
\label{del}
\end{figure}

\begin{figure}[tb]
\hfil
\includegraphics[width=40mm]{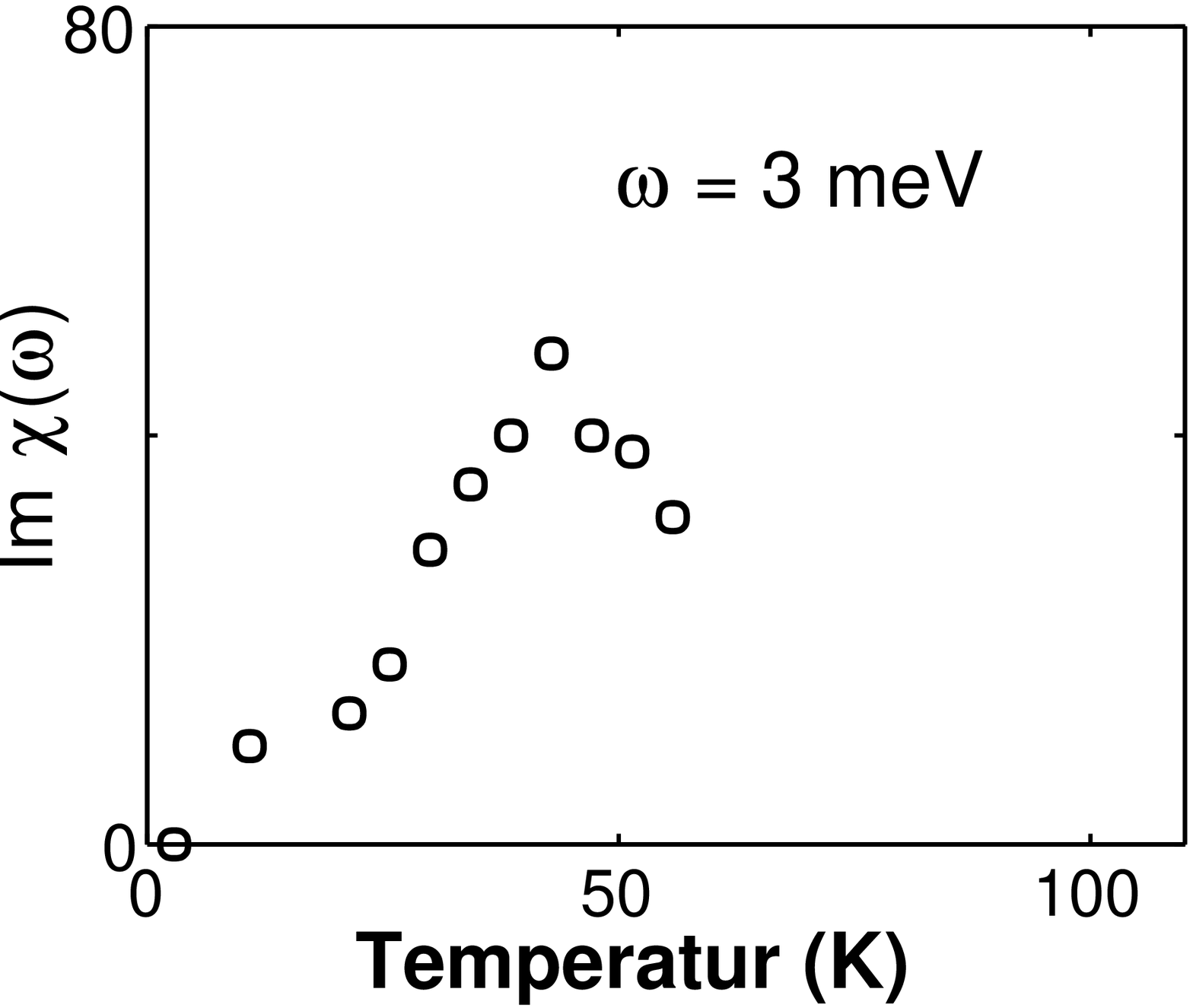}
\hfil
\includegraphics[width=40mm]{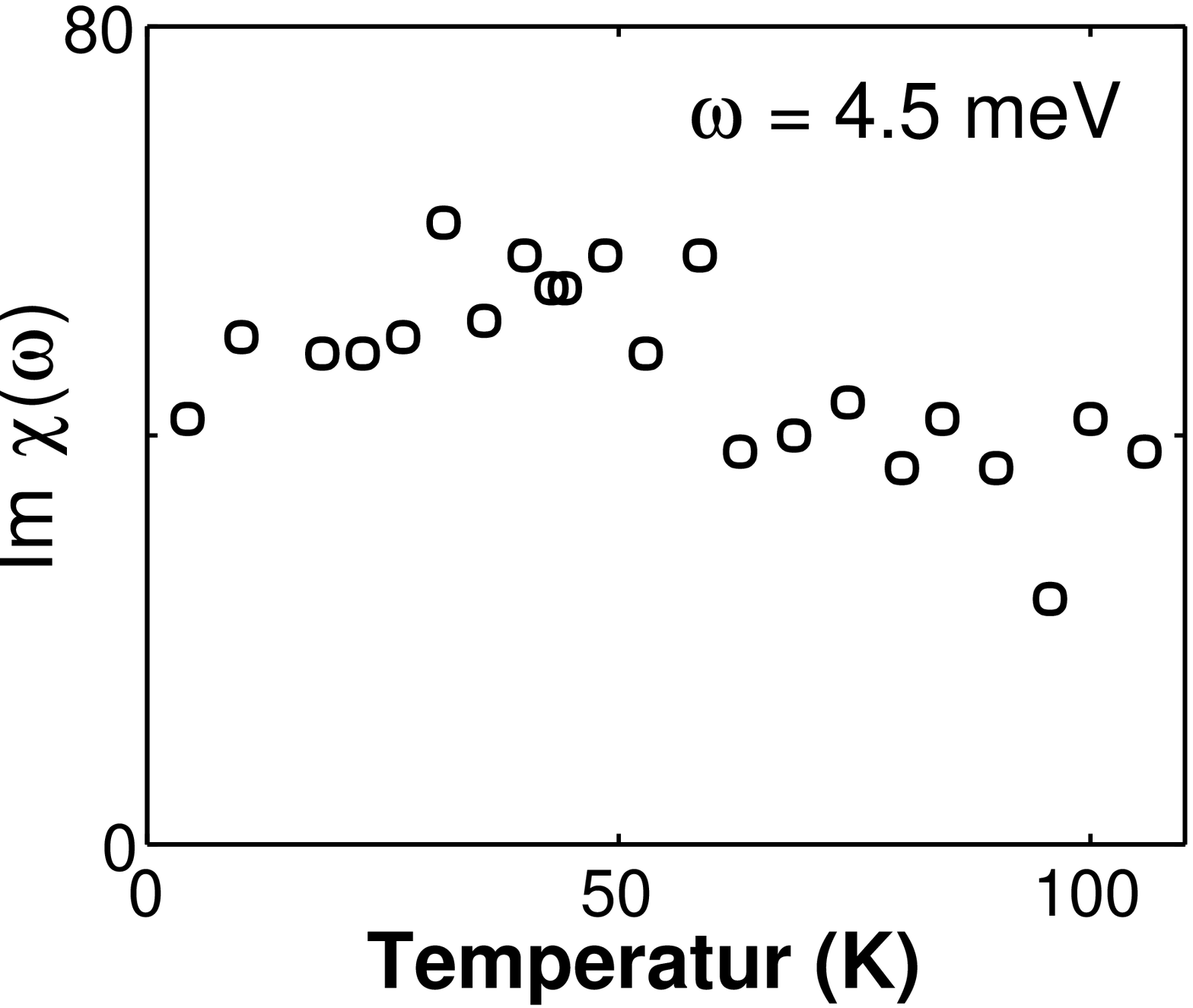}
\hfil
\caption{Temperature evolution of $\chi(\mathbf{q},\omega)$ at the 
incommensurate peak position for energies of 3 meV and 4.5 meV. 
Data is taken from Yamada {\it et al.}\cite{Yamada} whose sample 
was $La_{1.85}Sr_{0.15}C_uO_{4} (T_{c} = 37.3K)$. Notice the spectrum develops 
a gap for the 3 meV case on cooling below $T_{c}$
}
\label{yamadafig}
\end{figure}

To determine which scenario actually applies to real high $T_c$ samples, we
need to rely on experiments. In the following we would like to argue that
experiments  suggest the first scenario as most plausible in the high $T_c$
samples.  The appearance of Fermi arcs in the first scenario implies a small
energy scale $\Delta_{SCspin}$ for the spin excitations associated with the
superconducting state. (See Fig.[\ref{del}]) Such a small energy scale was
observed in the experiment of Ref. \onlinecite{Yamada} (see Fig.[\ref{yamadafig}]),
where it was found that the spin susceptibility $\chi(\omega)$ decreases below
$T_c$ only when $\omega < 4$meV. This indicates that $\Delta_{SCspin}$ is as
small as a few meV. Direct evidence for Fermi arcs comes from their
observation in a recent photoemission experiment.\cite{arc} The expected well
defined electron-like quasiparticles near the Fermi arcs were also observed.

\section{Conclusion and open questions}\label{Conclusion}

In this paper we have given an account of the spin correlations in the ASL
phase through the calculation of $O(1/N)$ contributions depicted in
Fig.[\ref{Pol1N}]. It was shown how the gauge fluctuations strongly enhance
the staggered spin correlations leaving the uniform spin correlation
unaffected. This result is very natural within this gauge fluctuation picture
where the uniform spin correlation is protected by current conservation and
cannot have any anomalous dimension.  From this perspective it is also easy to
account for the qualitative different behaviors of planar $C_u$ and $O$ spin
lattice relaxation rates seen in NMR experiments where $C_u$ probes the
enhanced staggered correlations of the ASL down to a temperature scale $T_{m}
\sim T_{T_{1}}$, where the $C_u$ $1/(T_{1}T)$ peaks, which we associate with the
appearance of an energy gap in the gauge spectrum.  What is really remarkable
in the ASL picture is that enhanced  staggered spin correlations are obtained
while at the same time having a small single particle density of states within
the pseudo-gap phase \emph{without any fine tuning}. As the doping $x$
decreases, the pseudo-gap increases and, surprisingly, the staggered spin
correlations also increase.  This strange experimental behavior is explained
naturally by the ASL picture.

It should be noted here that our ASL shows qualitative similarities with the
quantum critical point scenario of 
Refs. \onlinecite{Barzykin, CSY}.
However as was stressed in the paper,\cite{Barzykin} the appearance of the 
pseudo-gap
which they associate with the suppression of the spectral weight of spin waves
characteristic of the quantum disordered (QD) regime should affect the low
frequency dynamics for {\it both} $\mathbf{q} = 0$ and $\mathbf{q} =
(\pi,\pi)$. Within this framework it then seems hard to account for the
qualitative different behaviors seen at the $C_u$ and $O$ sites in NMR
experiments. As we have stressed many times within the ASL this difference is
protected by the $U(1)$ gauge structure. Also, in our ASL approach, we do not
assume, in contrast to the quantum-critical-point approach, any nearby
symmetry breaking phase and we do not require any strongly fluctuating order
parameters to give us critical behavior. The ASL can by itself appear as a
stable quantum phase, or as a phase transition point between two states with
new kinds of order - quantum order. The two states can have the \emph{same
symmetry} and no order parameters.\cite{Wqo} Hence the ASL associated with
the transition between quantum orders can show scaling properties divorced
from any critical point studied so far. 


{}From this perspective the ASL is just one of a whole slew of possible
quantum orders characterized by a massless $U(1)$ gauge field coupled to
massless Dirac fermions. It is also clear that the destruction of the ASL via
the transition into a phase with a massive $U(1)$ gauge field
(phenomenologically described above via the introduction of the mass-scale
$m$) demands a much more careful analysis and highlights the main difficulty
with the slave boson approach to the $tJ$-model. Within this scheme it
currently seems to be impossible or at least very challenging to describe this
energy/temperature regime around $m/T_{m}$ theoretically.  The reason for that
is related to the need of introducing a scale which is not tied to boson
condensation and hence divorced from the mean-field energies and the
corresponding degrees of freedom.  


Through our examination of the experimental data, we find that the ASL, plus
spin-charge recombination at lower energies, provides a consistent and natural
(with no fine tuning) description of underdoped cuprate superconductors. This
is the main result and the bottom line of this paper.

We would like to thank P. A. Lee and M. Kastner
for many helpful discussions. W.R. would particularly like to thank A. Seidel, for his mathematical insights.  This work is supported by NSF Grant No.
DMR--01--3156 and by NSF-MRSEC Grant No. DMR--98--08941.

\begin{widetext}
\appendix

\section{Spin operators in the continuum limit}
\label{appA}

In this appendix we derive the expression for the spin operators near
$\mathbf{q} = (0,0), (\pi,\pi)$ and $(\pi,0)$ in terms of the spinon field
$\Psi$ defined near the four nodes in momentum space.

Let us look at the staggered correlation in detail.
\begin{equation}\label{spinoperatorQ}
\vec{S}(\mathbf{Q}+\mathbf{q}) =
\frac{1}{2}\sum_{\mathbf{p}}f^{\dag}(\mathbf{p}-\mathbf{q})
\vec{\tau}f(\mathbf{p}+\mathbf{Q})
\end{equation}
where $\mathbf{Q} = (\pi/a,\pi/a)$ and we have suppressed the frequency
index.
After introducing
\begin{eqnarray}
f_{e} \equiv \frac{1}{\sqrt{2}}(f({\bf q}) + f({\bf q+Q})) \nonumber \\
f_{o} \equiv \frac{1}{\sqrt{2}}(f({\bf q}) - f({\bf q+Q}))
\end{eqnarray}
the expression for the staggered spin operator can be rewritten
\begin{equation}
\vec{S}(\mathbf{Q} +\mathbf{q}) =
\frac{1}{4}\sum_{\mathbf{p}}\left[f_{e}^{\dag}(\mathbf{p}-\mathbf{q}),
f_{o}^{\dag}(\mathbf{p}-\mathbf{q})\right]\vec{\tau}
\left(\sigma_{3}-i\sigma_{2}\right) {f_{e}(\mathbf{p}) 
\choose f_{o}(\mathbf{p})}
\end{equation}

Splitting the sum $\sum_{\mathbf{p}} =
\sum_{\mathbf{\tilde{Q}}+\mathbf{k}} +
\sum_{\mathbf{\tilde{\tilde{Q}}}+\mathbf{k'}}$ where $\mathbf{\tilde{Q}} =
(\pi/2,\pi/2) \quad \mathbf{\tilde{\tilde{Q}}} = (-\pi/2,-\pi/2)$ 
yields
\begin{eqnarray}
\vec{S}(\mathbf{Q}+\mathbf{q}) &=&
\frac{1}{4}\sum_{\mathbf{k}}\left[f_{e}^{\dag}
(\mathbf{k}-\mathbf{q}+\mathbf{\tilde{Q}}), f_{o}^{\dag}
(\mathbf{k}-\mathbf{q}+\mathbf{\tilde{Q}})\right]\vec{\tau}
\left(\sigma_{3}-i\sigma_{2}\right) {f_{e}(\mathbf{k}+\mathbf{\tilde{Q}})) 
\choose f_{o}(\mathbf{k}+\mathbf{\tilde{Q}})} \nonumber \\
&+&\frac{1}{4}\sum_{\mathbf{k'}}\left[f_{e}^{\dag}
(\mathbf{k'}-\mathbf{q}+\mathbf{\tilde{\tilde{Q}}}), f_{o}^{\dag}
(\mathbf{k'}-\mathbf{q}+\mathbf{\tilde{\tilde{Q}}})\right]
\vec{\tau}\left(\sigma_{3}-i\sigma_{2}\right) {f_{e}
(\mathbf{k'}+\mathbf{\tilde{\tilde{Q}}})) 
\choose f_{o}(\mathbf{k'}+\mathbf{\tilde{\tilde{Q}}})}
\end{eqnarray}
Noting $\mathbf{\tilde{\tilde{Q}}} = \mathbf{\tilde{Q}} + \mathbf{Q}$ and
using $f_{e}(\mathbf{p} +\mathbf{Q}) = f_{e}(\mathbf{p}) \quad
f_{o}(\mathbf{p} +\mathbf{Q}) = -f_{o}(\mathbf{p})$ we arrive at
\begin{eqnarray}
\vec{S}(\mathbf{Q}+\mathbf{q}) &=&
\frac{1}{2}\sum_{\mathbf{k}}\left[f_{e}^{\dag}
(\mathbf{k}-\mathbf{q}+\mathbf{\tilde{Q}}), f_{o}^{\dag}
(\mathbf{k}-\mathbf{q}+\mathbf{\tilde{Q}})\right]
\vec{\tau}\sigma_{3} {f_{e}(\mathbf{k}+\mathbf{\tilde{Q}})) 
\choose f_{o}(\mathbf{k}+\mathbf{\tilde{Q}})} \nonumber \\
&=& \frac{1}{2}\sum_{\mathbf{k}}
\bar{\Psi}_{1}(\mathbf{k}-\mathbf{q})\openone\vec{\tau}\Psi_{1}(\mathbf{k})
\end{eqnarray}
where $\bar{\Psi}_{1}(\mathbf{k}-\mathbf{q}) =
\left(f_{e}^{\dag}(\mathbf{k}-\mathbf{q}+\mathbf{\tilde{Q}}),
-f_{o}^{\dag}(\mathbf{k}-\mathbf{q}+\mathbf{\tilde{Q}})\right)$ and
$\mathbf{k}$ is measured away from the node of type
\newline
$1 \equiv \mathbf{\tilde{Q}} = \pm(\pi/2,\pi/2)$.  The remaining contribution
arises from the other two nodes in the spectrum at $\pm(\pi/2,-\pi/2)$.

Following the same calculation for the uniform correlation function 
\begin{equation}\label{spinoperator}
\vec{S}(\mathbf{q}) =
\frac{1}{2}\sum_{\mathbf{p}}f^{\dag}(\mathbf{p}-\mathbf{q})
\vec{\tau}f(\mathbf{p})
\end{equation}
we find 
\begin{equation}
\vec{S}(\mathbf{q}) =
\frac{1}{2}\sum_{\mathbf{k}}
\bar{\Psi}_{1}(\mathbf{k}-\mathbf{q})\sigma_{3}\vec{\tau}\Psi_{1}(\mathbf{k})
\end{equation}
with $\bar{\Psi}_{1}(\mathbf{k}-\mathbf{q}) =
\left(f_{e}^{\dag}(\mathbf{k}-\mathbf{q}+\mathbf{\tilde{Q}}),
-f_{o}^{\dag}(\mathbf{k}-\mathbf{q}+\mathbf{\tilde{Q}})\right)$ and
$\mathbf{k}$ is measured away from the node of type \newline $1 \equiv
\mathbf{\tilde{Q}} = \pm(\pi/2,\pi/2)$. As before the remaining contribution
arises from the other two nodes.

The correlation near $\mathbf{q} = (\pi,0)$ is obtained in a similar fashion
the only difference here is that for this momentum transfer the two types of
nodes get mixed and hence when we split up the sum over momentum space we need
to consider all four nodes simultaneously. Under this proviso the calculation
goes through as above and we end up with
\begin{equation}
\vec{S}(\mathbf{Q_x} + \mathbf{q}) =
\frac{1}{2}\sum_{\mathbf{k}}
\bar{\Psi}(\mathbf{k}-\mathbf{q})\pmatrix{0&\sigma_{1}\cr\sigma_{1}&0\cr}
\vec{\tau}\Psi(\mathbf{k})
\end{equation}
where
\begin{displaymath}
\bar{\Psi}(\mathbf{k}-\mathbf{q}) =
\left[f_{e}^{\dag}(\mathbf{k}-\mathbf{q}+\mathbf{Q_1}),
-f_{o}^{\dag}(\mathbf{k}-\mathbf{q}+\mathbf{Q_1}),-f_{o}^{\dag}
(\mathbf{k}-\mathbf{q}+\mathbf{Q_2}),f_{e}^{\dag}
(\mathbf{k}-\mathbf{q}+\mathbf{Q_2})\right]
\end{displaymath}
with $\mathbf{Q_1} = (\pi/2,\pi/2)$ and $\mathbf{Q_2} = (\pi/2, -\pi/2)$

\section{Two-loop calculation of spin correlation}
\label{appB}

We employ a four dimensional representation  of the Dirac algebra
$\{\gamma_{\mu},\gamma_{\nu}\}=2\delta_{\mu\nu}$ ($\mu,\nu = 0,1,2$) in the
main body of the paper ie $\Tr \openone = 4$.  To perform the $\Tr$ over the
spinor indices we need the following identity

\begin{displaymath}
\Tr \left[ \gamma^\epsilon \gamma^\delta \gamma^\mu \gamma^\beta \gamma^\nu
\gamma^\alpha \right ] = \delta^{\alpha\nu} \Tr\left[\gamma^\epsilon
\gamma^\delta \gamma^\mu \gamma^\beta \right ] -\delta^{\alpha\beta}
\Tr\left[\gamma^\epsilon \gamma^\delta \gamma^\mu \gamma^\nu \right ]
+\delta^{\alpha\mu}\left[\gamma^\epsilon \gamma^\delta \gamma^\beta
\gamma^\nu \right ] -\delta^{\alpha\delta}\left[\gamma^\epsilon \gamma^\mu
\gamma^\beta \gamma^\nu \right ]+\delta^{\alpha\epsilon}\left[\gamma^\delta
\gamma^\mu \gamma^\beta \gamma^\nu \right ]
\end{displaymath}
which can be simply derived from the Dirac algebra by commuting
$\gamma$-matrices through and using the cyclic property of the trace $\Tr$.
Using the above identity we can simplify

\begin{equation}
[\ref{Pol1N}(A)] = \int \frac{d^d k}{(2\pi)^d} \int \frac{d^d q}{(2\pi)^d}
\Tr \left [\frac{8}{N}
\frac{(p+k)_{\epsilon}\gamma^{\epsilon}k_{\delta}
\gamma^{\delta}\gamma^{\mu}\gamma^{\beta}(k+q)_{\beta}\gamma^{\nu}
\gamma^{\alpha}k_{\alpha}
(\delta_{\mu \nu} \vec{q}^2 - q_{\mu}q_{\nu})}{
(\vec{p}+\vec{k})^2 \vec{k}^2 (\vec{k}+\vec{q})^2 \vec{k}^2 |\vec{q}|^3} 
\right ] \\
\end{equation}
to
\begin{equation}\label{Aqk}
[\ref{Pol1N}(A)] = \frac{64}{N}\int \frac{d^d k}{(2\pi)^d} \int \frac{d^d
q}{(2\pi)^d} \frac{[\vec{k}^2(\vec{p}+\vec{k})\cdot \vec{q}
(\vec{k}+\vec{q})\cdot \vec{q} - 2(\vec{p}+\vec{k})\cdot \vec{k}
(\vec{k}+\vec{q})\cdot \vec{q}
\vec{k}\cdot\vec{q}]}{(\vec{p}+\vec{k})^2
(\vec{k}+\vec{q})^2\vec{k}^4\vec{q}^3} \nonumber
\end{equation}
and for diagram [\ref{Pol1N}(C)] we obtain
\begin{eqnarray}\label{Cqk}
[\ref{Pol1N}(C)] &=& \int \frac{d^d k}{(2\pi)^d} \int \frac{d^d q}{(2\pi)^d}
\Tr \left [\frac{8}{N}
\frac{(p+k)_{\epsilon}\gamma^{\epsilon}\gamma^{\mu}(p+k+q)_{\delta}
\gamma^{\delta}(k+q)_{\beta}\gamma^{\beta}
\gamma^{\nu}k_{\alpha}\gamma^{\alpha}(
\delta_{\mu \nu}\vec{q}^2 - q_{\mu}q_{\nu})}{
(\vec{p}+\vec{k})^2(\vec{p}+\vec{k}+\vec{q})^2
(\vec{k}+\vec{q})^2\vec{k}^2|\vec{q}|^3} \right ] \\
&=& \frac{64}{N} \int \frac{d^d k}{(2\pi)^d} \int \frac{d^d q}{(2\pi)^d}
\frac{
F(\vec{q},\vec{k},\vec{p})}{
(\vec{p}+\vec{k})^2(\vec{p}+\vec{k}+\vec{q})^2
(\vec{k}+\vec{q})^2\vec{k}^2|\vec{q}|^3}
\end{eqnarray}
where
\begin{eqnarray*}
F(\vec{q},\vec{k},\vec{p}) = \Bigg [ & &\vec{k}^2 \big ( \vec{q}^2\vec{k}^2
- \frac{1}{2} \vec{k}\cdot\vec{q}\vec{p}^2 + 2\vec{k}\cdot\vec{q}\vec{q}^2 +
\vec{k}\cdot\vec{p}\vec{q}\cdot\vec{p} - \frac{1}{4}\vec{q}^2\vec{p}^2 +
\vec{q}^4 + (\vec{q}\cdot\vec{p})^2 \big ) \\
&+&  \frac{1}{2}(\vec{k}+\vec{q})^2\big (2\vec{q}^2\vec{k}\cdot\vec{p} +
\vec{p}^2\vec{k}\cdot\vec{q} - 2\vec{k}\cdot\vec{p}\vec{q}\cdot\vec{p} -
\frac{1}{2} \vec{q}^2\vec{p}^2 \big ) \\
&+&  (\vec{k} +\vec{p})^2 \big(\vec{q}^2\vec{p}\cdot\vec{q} + \vec{q}^2
\vec{k}\cdot\vec{p} - \frac{1}{2}\vec{q}^2\vec{p}^2 \big ) \\
&+&  \frac{1}{4}\vec{q}^4\vec{p}^2 - \vec{p}^2\vec{q}^2\vec{q}\cdot\vec{p} +
\frac{1}{2}\vec{q}^2\vec{p}^4 \Bigg]
\end{eqnarray*}

In both expressions (\ref{Aqk},\ref{Cqk}) the integration over $\vec{k}$ is
convergent in $d = 3$ and can be performed noting that
\begin{eqnarray}
\int \frac{d^3 k}{(2\pi)^3}
\frac{1}{\vec{k}^2(\vec{k}+\vec{p})^2(\vec{k}+\vec{q})^2} &=& \frac{1}{8
|\vec{p}| |\vec{q}| |\vec{q}-\vec{p}|} \\
\int \frac{d^3 k}{(2\pi)^3}
\frac{1}{\vec{k}^2(\vec{k}+\vec{p})^2(\vec{k}+\vec{q})^2
(\vec{k}+\vec{p}+\vec{q})^2} &=& \frac{1}{8 |\vec{q}||\vec{p}|
\vec{q}\cdot\vec{p}}\left[ \frac{1}{|\vec{p}-\vec{q}|} 
-  \frac{1}{|\vec{q}+\vec{p}|} \right ]
\end{eqnarray}

Thus after the  $\vec{k}$ integration we arrive at
\begin{equation}\label{Aq}
[\ref{Pol1N}(A)] =\frac{2}{N}\int \frac{d^d q}{(2\pi)^d} \left [
\frac{2\vec{p}\cdot\vec{q} + \vec{p}^2}{|\vec{q}|^3 |\vec{p} + \vec{q}|} -
\frac{\vec{p}\cdot\vec{q}}{|\vec{p}|\vec{q}^2|\vec{p}+\vec{q}|} -
\frac{|\vec{p}|}{|\vec{q}|^3}\right]
\end{equation}
\begin{equation} \label{Cq}
[\ref{Pol1N}(C)] = \frac{4}{N}\int \frac{d^d q}{(2\pi)^d}\bigg
[\frac{2}{\vec{q}^2} + \frac{|\vec{p}|}{|\vec{q}|^3} - \frac{\vec{p}^2 +
2\vec{p}\cdot\vec{q}}{|\vec{q}|^3|\vec{p}+\vec{q}|} -\frac{4
|\vec{p}|}{\vec{q}^2|\vec{p}+\vec{q}|} - \frac{\vec{q}^2\vec{p}^2 + 2
\vec{p}^4}{\vec{p}\cdot\vec{q}\vec{q}^2|\vec{p}||\vec{p}+\vec{q}|} \bigg ]
\end{equation}

Adding the contributions
\begin{equation}\label{2A+C}
2\times[\ref{Pol1N}(A)] + [\ref{Pol1N}(C)] = \frac{4}{N}\int \frac{d^d
q}{(2\pi)^d}\left[\frac{2}{\vec{q}^2}  -\frac{4
|\vec{p}|}{\vec{q}^2|\vec{p}+\vec{q}|}-
\frac{\vec{p}\cdot\vec{q}}{|\vec{p}|\vec{q}^2|\vec{p}+\vec{q}|} -
\frac{\vec{q}^2\vec{p}^2 + 2
\vec{p}^4}{\vec{p}\cdot\vec{q}\vec{q}^2|\vec{p}||\vec{p}+\vec{q}|} \right]
\end{equation}

As noted previously because of the $\vec{p}\cdot\vec{q}$ term in the denominator of
the last term above we wont use dimensional regularization. Therefore we set
$d=3$ and introduce an upper cut-off $\Lambda$ to regularize the above
integrals.  Also note that we will neglect the first term in (\ref{2A+C})
which is linearly divergent whose appearance is tight to the regularization
via a momentum cut-off ( such
divergences do not appear in Lorentz invariant regularization schemes).

Under this proviso we obtain the following result for the last three terms in
(\ref{2A+C})
\begin{equation}
\frac{4}{N}\left[ -\frac{8|\vec{p}|}{4\pi^2} -
\frac{4|\vec{p}|}{4\pi^2}ln\frac{\Lambda^2}{\vec{p}^2} +
\frac{2|\vec{p}|}{36\pi^2} +
\frac{|\vec{p}|}{12\pi^2}ln\frac{\Lambda^2}{\vec{p}^2} +
\frac{|\vec{p}|}{8\pi^2}(C_{1} + C_{2}) +
\frac{2|\vec{p}|}{8\pi^2}ln\frac{\Lambda^2}{\vec{p}^2} \right]
\end{equation}
where the last term (\ref{2A+C}) is given by
\begin{eqnarray}
\int \frac{d^3 q}{(2\pi)^3}\frac{\vec{q}^2\vec{p}^2 + 2
\vec{p}^4}{\vec{p}\cdot\vec{q}\vec{q}^2|\vec{p}||\vec{p}+\vec{q}|} &=&
\frac{|\vec{p}|}{8\pi^2}\left[ \int_{0}^{1} dy
\frac{2+y}{y\sqrt{1+y}}ln\frac{\sqrt{1+y}-\sqrt{y}}{\sqrt{1+y}+\sqrt{y}} +
\int_{1}^{\frac{\Lambda^2}{\vec{p}^2}} dy
\frac{2+y}{y\sqrt{1+y}}ln\frac{\sqrt{1+y}-1}{\sqrt{1+y}+1} \right]\\
&=& -\frac{|\vec{p}|}{8\pi^2}\left( C_{1} + C_{2} +
2ln\frac{\Lambda^2}{\vec{p}^2}\right)
\end{eqnarray}
and
\begin{eqnarray*}
C_{1} &=& -\int_{0}^{1} dy
\frac{2+y}{y\sqrt{1+y}}ln\frac{\sqrt{1+y}-\sqrt{y}}{\sqrt{1+y}+\sqrt{y}} =
7.748128723 \\
C_{2} &=&
4ln(\sqrt{2}-1)+4+ln(3-2\sqrt{2})^2-2ln(3-2\sqrt{2})
+2ln(3-2\sqrt{2})\sqrt{2} =  2.121475665
\end{eqnarray*}

Finally extracting the logarithmically divergent terms we arrive at the
result stated in the body of the text.

\section{Analytic continuation of $1/N$ correction}
\label{appC}

Having obtained the $1/N$ correction we would now like to reexponentiate our
result in the form
\begin{equation}\label{Naivesum}
\langle S^{+}_{s}(\vec{q}) S^{-}_{s}(\vec{q}) \rangle_{0} = -
\frac{\sqrt{\vec{q}^2}}{16} -\frac{8}{12 \pi^2 N}\sqrt{\vec{q}^2} 
\ln\bigg(\frac{\Lambda^2}{\vec{q}^2}\bigg) \sim 
- \frac{\Lambda^{2\nu}}{16}\left(q^2\right)^{1/2-\nu}  \quad \nu 
= \frac{32}{3\pi^2N}
\end{equation}

This however immediately confronts us with the problem that after analytically
continuing (\ref{Naivesum}) the susceptibility has the wrong sign for $2\nu >
1$. It is however hard to understand why the spin operator cannot have an
anomalous dimension bigger than $\frac{1}{2}$.  In order to analyze this issue
it is helpful to look at the corresponding Euclidean real space correlations.
Let's take the  case of a general staggered spin correlation of the form
\begin{equation}
\langle S_s^{+}(\vec{x})S_s^{-}(0)\rangle = 
\frac{1}{16\pi^2}(-1)^{\mathbf{x}}\frac{C}{|\vec{x}|^{4-2\nu}}
\end{equation}
where the spin operator has anomalous dimension $\nu$. We now Fourier
transform this to obtain
\begin{equation}
\frac{1}{4\pi} |\vec{q}|^{1-2\nu} \int_0^\infty \frac{du}{u^{3-2\nu}}sin(u)
\end{equation}
For $ 0 < |Re(2\nu-2)|<1 $ this can be evaluated \cite{GR} to give
\begin{equation} 
\frac{1}{4\pi} \Gamma(2\nu-2)sin((\nu-1)\pi)|\vec{q}|^{1-2\nu}
\end{equation}

This leads after analytic continuation to the result equation(\ref{sspin}). From here
it is easy to see how the $\Gamma$-function and the $sin$ which are missed in
the naive exponentiation of the Euclidean momentum space result conspire to
give the correct sign for the spin correlation no matter what the size of the
anomalous dimension. In particular we can continue the above for the case $\nu
\rightarrow 0$ to give the correct mean-field result.

\section{Two-loop calculation in the massive phase of the $U(1)$ gauge
field}
\label{appD}

This appendix gives some details of the calculation leading to the $O(1/N)$
corrections to the staggered spin correlation in the case when the gauge field
is in the massive phase with propagator
\begin{displaymath}
D_{\mu \nu}(\vec{q}) = \frac{8}{N \sqrt{\vec{q}^2 + m^2}}\big (
\delta_{\mu\nu} - \frac{q_{\mu}q_{\nu}}{\vec{q}^2} \big )
\end{displaymath}
We can use the results derived in Appendix A for the integrals over $\vec{k}$
(\ref{2A+C}) since they are unaffected by the change in the gauge propagator.
\begin{equation}\label{2A+Cm}
2[\ref{Pol1N}(A)] + [\ref{Pol1N}(C)] = \frac{4}{N}\int \frac{d^d
q}{(2\pi)^d}\left[-\frac{4
|\vec{p}|}{|\vec{q}|\sqrt{\vec{q}^2+m^2}|\vec{p}+\vec{q}|} -
\frac{\vec{p}\cdot\vec{q}}{|\vec{p}||\vec{q}|
\sqrt{\vec{q}^2+m^2}|\vec{p}+\vec{q}|} 
- \frac{\vec{q}^2\vec{p}^2 + 2 \vec{p}^4}{
\vec{p}\cdot\vec{q}|\vec{q}|\sqrt{\vec{q}^2+m^2}|\vec{p}||\vec{p}
+\vec{q}|} \right]
\end{equation}
where we have neglected the linearly divergent term as discussed above.
After integration over $\vec{q}$ in $d=3$ we arrive at
\begin{eqnarray}\label{1N_m}
\frac{4}{N} \Bigg[ &-&\frac{2}{\pi^2}\left[\sqrt{\vec{p}^2+m^2}-m\right]
- \frac{2|\vec{p}|}{\pi^2}ln\frac{2\Lambda}{|\vec{p}| +
\sqrt{\vec{p}^2+m^2}} + \frac{\sqrt{\vec{p}^2+m^2}}{6\pi^2} -
\frac{1}{9\pi^2\vec{p}^2}\left[(\vec{p}^2+m^2)^{3/2} -m^3\right] +  \\
 &+&\frac{|\vec{p}|}{6\pi^2}ln\frac{2\Lambda}{|\vec{p}| +
\sqrt{\vec{p}^2+m^2}} - \frac{|\vec{p}|}{8\pi^2}\int_{0}^{1} dy
\frac{2+y}{\sqrt{y}\sqrt{y+(\frac{m}{|\vec{p}|})^2}
\sqrt{1+y}}ln\frac{\sqrt{1+y}-\sqrt{y}}{\sqrt{1+y}+\sqrt{y}} -  
\nonumber \\ \nonumber
 &-& \frac{|\vec{p}|}{8\pi^2} \int_{1}^{\frac{\Lambda^2}{\vec{p}^2}} dy
\frac{2+y}{\sqrt{y}\sqrt{y+(\frac{m}{|\vec{p}|})^2}
\sqrt{1+y}}ln\frac{\sqrt{1+y}-1}{\sqrt{1+y}+1} \Bigg]
\end{eqnarray}
It is not hard to check that in the limit $\frac{|\vec{p}|}{m} \rightarrow 0$
the terms proportional to $m$ cancel. In extracting the logarithmic term we
have to take a closer look at the last term which can't be evaluated in a
closed form.
\begin{figure}[tb]
\begin{center}
\includegraphics[width=70mm]{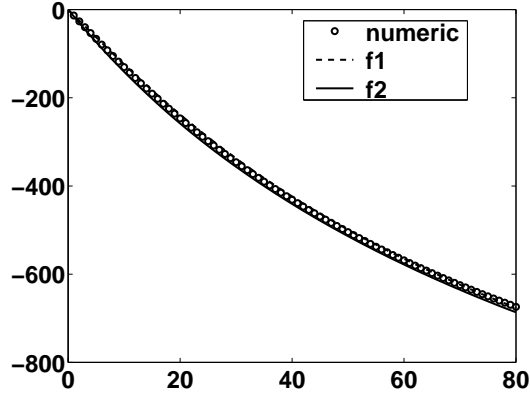}
\caption{Comparison between $f1 =
-2|\vec{p}|ln\frac{\Lambda^2}{\vec{p}^2+m^2}$,$f2 =
-4|\vec{p}|ln\frac{2\Lambda}{|\vec{p}|+\sqrt{\vec{p}^2+m^2}}$ and the
numeric evaluation of (\ref{numeric}) before analytic continuation
($m=30$).}
\label{comp1}
\end{center}
\end{figure}
In Fig.(\ref{comp1}) we compare two different functional forms with the
numeric evaluation of
\begin{equation}\label{numeric}
 |\vec{p}| \int_{1}^{\frac{\Lambda^2}{\vec{p}^2}} dy
\frac{2+y}{\sqrt{y}\sqrt{y+(\frac{m}{|\vec{p}|})^2}
\sqrt{1+y}}ln\frac{\sqrt{1+y}-1}{\sqrt{1+y}+1}
\end{equation}
Note that there is virtually no difference between $f1 =
-2|\vec{p}|ln\frac{\Lambda^2}{\vec{p}^2+m^2}$ and $f2 =
-4|\vec{p}|ln\frac{2\Lambda}{|\vec{p}|+\sqrt{\vec{p}^2+m^2}}$. However this
comparison has to be taken with a grain of salt as we are still in Euclidean
space but are interested in the analytically continued forms.
\begin{figure}[tb]
\begin{center}
\includegraphics[width=70mm]{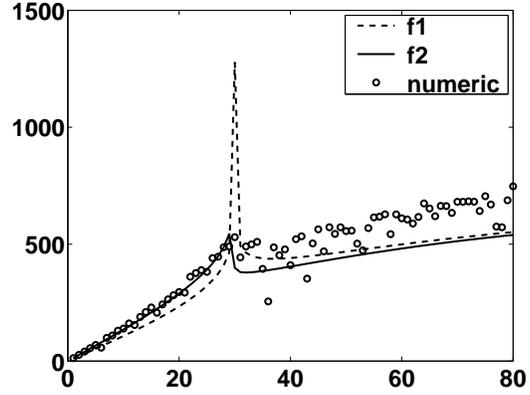}
\caption{Comparison between the analytically continued forms of $f1$, $f2$
and the numeric evaluation of (\ref{numeric}) after analytic continuation
($m=30$).}
\label{comp2}
\end{center}
\end{figure}
In Fig.(\ref{comp2}) we plot the analytically continued ($ip_{0}\rightarrow
\omega + i\epsilon$) forms of the above functions. From this comparison it is
now obvious that $f1$ which has developed a  pole at $m$ is unsuitable as an
approximation and we take $f2$ which fits the numerically integrated form
quite well.  Collecting the $ln$-terms in (\ref{1N_m}) we obtain
\begin{equation}
2\times[\ref{Pol1N}(A)]_{m} + [\ref{Pol1N}(C)]_{m} =
-\frac{16|\vec{q}|}{3\pi^2N}ln\frac{2\Lambda}{|\vec{q}|
+\sqrt{\vec{q}^2+m^2}}
\end{equation}

Analytically continuing and combining with the mean-field result
(\ref{Meanfield}) gives (\ref{resultm}).

\section{Low energy effective theory for the superconducting state}
\label{appE}

Starting from the mean-field Hamiltonian for the spinons
\begin{equation}
H = \frac{1}{2}\sum_{i,j,\sigma}\psi^{\dag}_{\sigma
i}\tilde{J}U_{ij}\psi_{\sigma j} 
 + \sum_{il}a^{l}_{0i}\frac{1}{2}\psi^{\dag}_{\sigma
i}\tau^{l}\psi_{\sigma i} 
\end{equation}
in the d-wave paired state
\begin{eqnarray}\label{d-wave}
U_{i,i+\hat{x}} &=& -\tau^{3}\chi + \tau^{1}\Delta \nonumber \\
U_{i,i+\hat{y}} &=& -\tau^{3}\chi - \tau^{1}\Delta \nonumber \\
a_{0}^{3} &\neq& 0 
\end{eqnarray} 
we obtain after Fourier transformation the following Lagrangian
\begin{eqnarray}
L &=& \sum_{\mathbf{q},\omega} \left[ f^{\dagger}_{\uparrow}(\mathbf{q},\omega), f_{\downarrow}(-\mathbf{q},-\omega)\right] \left( \begin{array}{l r}
-i\omega + \epsilon(\mathbf{q}) + a_{0}^3 & -\eta(\mathbf{q}) \\
-\eta(\mathbf{q}) & -i\omega -\epsilon(\mathbf{q}) - a_{0}^3 \end{array}\right) { f_{\uparrow}(\mathbf{q},\omega) \choose f^{\dagger}_{\downarrow}(-\mathbf{q},-\omega)} \\
\epsilon(\mathbf{q}) &=& -2\tilde{J}\chi(cos(q_{x}a)+cos(q_{y}a)) \quad
\eta(\mathbf{q}) = -2\tilde{J}\Delta (cos(q_{x}a)-cos(q_{y}a)) 
\end{eqnarray}
which results with
\begin{eqnarray}
S^{+}(\mathbf{q},\omega) = \frac{1}{2}\sum_{\mathbf{k},\omega_{k}}f^{\dagger}_{\uparrow}(\mathbf{k},\omega_{k})f_{\downarrow}(\mathbf{k+q},\omega_{k} + \omega_{q}) \nonumber \\
S^{-}(\mathbf{q},\omega) = \frac{1}{2}\sum_{\mathbf{k},\omega_{k}}f^{\dagger}_{\downarrow}(\mathbf{k},\omega_{k})f_{\uparrow}(\mathbf{k+q},\omega_{k} + \omega_{q})
\end{eqnarray}
in
\begin{equation}
\langle S^{+}(\mathbf{q},\omega) S^{-}(-\mathbf{q},-\omega) \rangle = 
\frac{1}{4}\sum_{\mathbf{k},\omega_{k}} \frac{\left( -i\omega_{k} - \epsilon(\mathbf{k}) - a_{0}^3 \right) \left( i\omega_{k} - i\omega_{q} + \epsilon(\mathbf{k-q}) + a_{0}^3 \right) - \eta(\mathbf{k})\eta(\mathbf{k-q})}{\left[ \omega_{k}^2 + \left(\epsilon(\mathbf{k}) + a_{0}^3\right)^2 + \eta(\mathbf{k})^2\right] \left[ \left(\omega_{k}-\omega_{q}\right)^2 + \left(\epsilon(\mathbf{k-q}) + a_{0}^3 \right)^2 + \eta(\mathbf{k-q})^2 \right] }
\end{equation}
next we split  $\sum_{\mathbf{k}} = \sum_{\mathbf{Q_{i} + \tilde{p}}}$ where
$\mathbf{Q_{i}}$ with $i = 1 ...4$ corresponds to $(\pi/2,\pi/2),
(-\pi/2,-\pi/2), (\pi/2,-\pi/2), (-\pi/2, \pi/2)$ in this order.  To extract
the correlations near $\mathbf{Q_{AF}} = (\pi,\pi)$ we write $\mathbf{q} =
\mathbf{Q_{AF}} + \tilde{k}$

Concentrating on one of the terms in the sum - e.g. about $\mathbf{Q}_{1} =
(\pi/2,\pi/2)$ we can expand
\begin{equation}
\epsilon(\mathbf{Q_{1} + \tilde{p}}) \sim v_{f} p_{1} \quad \eta(\mathbf{Q_{1} + \tilde{p}}) \sim -v_{2} p_{2} \quad
v_{f} \equiv 2\sqrt{2}\chi J \quad v_{2} \equiv 2\sqrt{2}\delta J \quad p_{1} \equiv \frac{\tilde{p}_{x} + \tilde{p}_{y}}{\sqrt{2}} \quad p_{2} \equiv \frac{-\tilde{p}_{x} + \tilde{p}_{y}}{\sqrt{2}} 
\end{equation}
to obtain
\begin{eqnarray}\label{expansion}
&&\langle S^{+}(\mathbf{Q_{AF} + \tilde{k}},\omega) S^{-}(-\mathbf{Q_{AF} - \tilde{k}},-\omega) \rangle \nonumber \\
&=&\frac{1}{4}\sum_{\mathbf{Q_{1}},\omega_{p}} \frac{\left( -i\omega_{p} - v_{f}p_{1} - a_{0}^3 \right) \left( i\omega_{p} - i\omega_{k} -v_{f}(p_{1} - k_{1}) + a_{0}^3 \right) + v_2 p_2 v_2 (p_2 - k_2)}{\left[ \omega_{p}^2 + \left(v_{f}p_1 + a_{0}^3\right)^2 + v_2^2p_2^2\right] \left[ \left(\omega_{p}-\omega_{k}\right)^2 + \left(-v_{f}(p_1-k_1) + a_{0}^3 \right)^2 + v_2^2(p_2 - k_2)^2 \right] } \nonumber \\
&+& \frac{1}{4}\sum_{\mathbf{Q_{i}},\omega_{p}, i = 2\ldots 4} \cdots
\end{eqnarray}
Note  $k_1 \equiv \frac{\tilde{k}_{x} + \tilde{k}_{y}}{\sqrt{2}} \quad k_2
\equiv \frac{-\tilde{k}_{x} + \tilde{k}_{y}}{\sqrt{2}}$ Now we approximate the
sums by integrals and define
\begin{eqnarray}
\bar{p}_1 \equiv v_{f}p_1 + a_0^3 \quad \bar{p}_2 \equiv v_2 p_2 \nonumber \\
\bar{k}_1 \equiv v_{f}k_1 + 2a_0^3 \quad \bar{k}_2 \equiv v_2 k_2 \nonumber
\end{eqnarray}
which allows us to rewrite the first term in (\ref{expansion})
\begin{equation}
\frac{1}{4} \int \frac{d^3\bar{p}}{(2\pi)^3v_{f}v_{2}} \frac{\vec{\bar{p}}\cdot\left(\vec{\bar{p}}-\vec{\bar{k}}\right) + i\bar{p}_1\omega_{\bar{k}} -i \omega_{\bar{p}}\bar{k}_1}{\left[ \vec{\bar{p}}^2 \left(\vec{\bar{p}}-\vec{\bar{k}}\right)^2\right]}
\end{equation}
Using the Feynman trick or otherwise it is not hard to convince oneself that
the last two terms cancel each other and the final result is
\begin{equation}  
-\frac{1}{64 v_{f} v_{2}}\sqrt{\omega_{k}^2 + \bar{k}_1^2 + \bar{k}_2^2}
\end{equation}
Combining all four contributions from the four nodes in this way and
translating the $\bar{k}$'s back to the $\tilde{k}$'s which measure momentum
along the ab axis away from $\mathbf{Q_{AF}}$ we obtain after analytic
continuation  
\begin{eqnarray}
&& \Im\langle S^{+}(\omega,\mathbf{Q_{AF} + \tilde{k}}) S^{-}(-\omega,-\mathbf{Q_{AF}-\tilde{k}}) \rangle \nonumber
\\
&=& \frac{1}{64 v_{f}v_{2}} \left\{\theta\left[ \omega^2 - (v_f k_1 +2 a_0^3)^2 - v_2^2k_2^2 \right]  \sqrt{\omega^2 - (v_f k_1 +2 a_0^3)^2 - v_2^2k_2^2} +\right.\nonumber \\
&& \quad \quad \quad  \theta\left[ \omega^2 - (v_f k_1 - 2 a_0^3)^2 - v_2^2k_2^2 \right]  \sqrt{\omega^2 - (v_f k_1 - 2 a_0^3)^2 - v_2^2k_2^2} + \nonumber \\
&& \quad \quad \quad \theta\left[ \omega^2 - (v_f k_2 +2 a_0^3)^2 - v_2^2k_1^2 \right]  \sqrt{\omega^2 - (v_f k_2 +2 a_0^3)^2 - v_2^2k_1^2} + \nonumber \\
 && \left. \quad \quad \quad \theta\left[ \omega^2 - (v_f k_2 - 2 a_0^3)^2 - v_2^2k_1^2 \right]  \sqrt{\omega^2 - (v_f k_2 - 2 a_0^3)^2 - v_2^2k_1^2}\right\} \\
k_1 &\equiv& \frac{\tilde{k}_x+\tilde{k}_y}{\sqrt{2}} \quad
k_2 \equiv \frac{-\tilde{k}_x +\tilde{k}_y}{\sqrt{2}}\quad
v_{f} \equiv 2\sqrt{2}aJ\chi \quad
v_{2} \equiv 2\sqrt{2}aJ\Delta \nonumber
\end{eqnarray}
the result stated in the main body of the paper.
\end{widetext}

\end{document}